\documentclass[11pt]{article}
\usepackage{jheppub}                

\title{The finite 't Hooft coupling correction on the jet quenching parameter
in a $\mathcal N=4$ super Yang-Mills plasma}
\author[\dag a]{Zi-qiang~Zhang,}
\author[\dag a]{De-fu~Hou,}
\author[\dag b,\dag a]{Hai-cang Ren,}
\affiliation[\dag a]{Institute of Particle Physics,Key Laboratory of QLP of  MOE, Huazhong Normal
University, Wuhan 430079, China }
 \affiliation[\dag b]{Physics Department, The
Rockefeller University, 1230 York Avenue, New York, NY 10021--6399}
\emailAdd{zhangzq@iopp.ccnu.edu.cn}\emailAdd{hdf@iopp.ccnu.edu.cn}\emailAdd{ren@mail.rockefeller.edu}
\abstract{We derive the quadratic action of the fluctuations around the classical world sheet underlying the jet quenching  from AdS/CFT. After obtaining the correspondence partition function, the expansion of the jet quenching parameter of $\mathcal N=4$ super symmetric Yang-Mills theory is carried out to the sub-leading term in the large 't Hooft coupling $\lambda$  at a nonzero temperature. The strong coupling corresponds to the semi-classical expansion of the string-sigma model, the gravity dual of the Wilson loop operator, with the sub-leading term expressed in terms of functional determinants of fluctuations. The contribution of these determinants are evaluated numerically. We find  the jet quenching parameter is reduced due to  world sheet fluctuations  by a factor $(1-1.97\lambda^{-1/2}) $. }

\begin{document}


\maketitle

\section{Introduction}

One of the main purposes of the heavy-ion collision experiments is to explore the QCD phase diagram and the properties of
new state of matter created through collisions at high energy density. The experimental program at the Relativistic Heavy Ion
Collider (RHIC) in Brookhaven National Laboratory has led to strong evidences that this new state of matter is in the
deconfinement phase of QCD, but behaves very differently from a free gas of partons. The name "strongly coupled quark-gluon
plasma (SQGP)" was given to this matter because of several striking experimental discoveries. Firstly, the measured elliptic
flow data can be well described by hydrodynamical models only with very low shear viscosity, suggesting that SQGP resembles a
nearly perfect liquid. Secondly, due to the interaction with the medium, high energy partons traversing the medium are strongly
quenched. This phenomenon is usually characterized by the so-called jet quenching parameter (or transport coefficient)
$\hat q$, which describes the average transverse momentum square transferred from the traversing parton, per unit mean free
path \cite{RB,XN}.
There are still some model-dependent uncertainties on the extracted values data of the transport coefficient from heavy-ion collisions, ranging  from $1\to 25GeV^2/fm$  \cite{Jorge,JD} , which is considerably larger
than those from perturbative QCD estimation \cite{XN1,GYQ}, higher values demand additional non-perturbative mechanisms.
Therefore, it is of great importance to get further information on the possible values of $\hat q $ in the strong coupling limit. The conventional nonperturbative first principle tool namely lattice QCD, cannot be applied for this purpose, which requires real-time dynamics.

A prominent implication of the AdS/CFT duality
\cite{Maldacena:1997re,Gubser:1998bc,Witten:1998qj,MadalcenaReview} is the correspondence between
the type IIB superstring theory formulated on AdS$_5\times S^5$ and $\mathcal N=4$ supersymmetric
Yang-Mills theory (SYM) in four dimensions. In particular, the supergravity limit of the string theory
corresponds to the leading behavior of SYM at large $N_c$ and large 't Hooft coupling
$\lambda \equiv g_{\rm YM}^2N_c = \frac{L^4}{\alpha'^2}$ with $L$ the AdS radius and $\alpha'$ the
reciprocal of the string tension. This relation thereby provides a powerful tool to explore the strongly
coupled QGP created in RHIC in spite of its underlying dynamics, QCD, is different from $\mathcal N=4$ SYM.
It is expected that some of the properties of
the latter is universal for all strongly interacting system. Computationally, AdS/CFT correspondence
translates the quantum field theory of supersymmetric Yang-Mills theories at strong 't Hooft
coupling and large number of colors $N_c$ into a classical field theory in a gravitational
background. The AdS/CFT correspondence has been successfully applied into the RHIC physics, including
the thermodynamical the transport properties of sQGP. One of the celebrated finding is the universal value
of the ratio between the shear viscosity and the entropy density, $\eta/s=\frac{1}{4\pi}$, for quantum field
theories admitting a holographic description. It was further conjectured that $\eta/s\ge\frac{1}{4\pi}$
for all matters. See \cite{SonReview} for a review on the subject and the references therein. The RHIC data of the viscosity ratio is compatible with this
lower bound. In addition, the entropy density of a $\mathcal N=4$ SYM plasma normalized by its
Stefan-Boltzmann limit equals to 3/4 \cite{GKP}, which is close to the corresponding quantity extracted from lattice QCD.

Motivated by these similarities,
Liu, Rajagopal and Wiedemann(LRW) computed the jet-quenching parameter of a $\mathcal N=4$
SYM plasma with the aid of AdS/CFT correspondence. The Eikonal approximation relates the jet quenching parameter
with the expectation value of an adjoint Wilson loop $W^A[{\cal C}]$ with ${\cal C}$ a rectangular contour of size
$L\times L_-$, where the sides with length $L_-$ run along the light-cone and the limit $L_-\to\infty$ is taken in the end
\cite{B. G. Zakharov}. Under the dipole approximation, which is valid for small transverse separation $L$,
the jet-quenching parameter $\hat{q}$ defined in Ref.\cite{RB} is extracted from the asymptotic expression for $TL<<1$
\begin{equation}
<W^A[{\cal C}]> \approx W_0\exp [-\frac{1}{4\sqrt{2}}\hat{q}L_-L^2]
\label{jet}
\end{equation}
with $T$ the temperature and the prefactor $W_0$ {\it independent} of $L$. The correspondence principle relates $<W^A[{\cal C}]>$ to the
quantum effective action of a type IIB superstring in $AdS_5\times S^5$ with a black hole.
In the strong coupling
limit, the action is proportional to the minimum area of the string world sheet in the $AdS_5$ bulk spanned
by the $C$ at the boundary. They find that \cite{liu}
\begin{equation}
\hat q_{SYM}^{(0)}=\frac{\pi^{3/2}\Gamma{(\frac{3}{4})}}{\Gamma{(\frac{5}{4})}}\sqrt{\lambda}T^3
\label{leading}
\end{equation}
Interestingly the magnitude of $\hat q_{SYM}^{(0)}$ turns out to be closer to the value extracted from RHIC data
\cite{K.J,A.D} than pQCD result for the typical value of the 't Hooft coupling,
$\lambda\simeq 6\pi$, of QCD. This proposal has attracted lots of interest. Alternately, the energy loss and
jet quenching problem was also studied from a drag force on the brane by several authors \cite{C.P,JF,KB,GB,Xiao}.
Their extracted jet quenching parameter is also proportional to $\sqrt{\lambda}T^3$ but with a smaller
coefficient (by $20$ percent).

The LRW formula is strictly valid only when the 't Hooft coupling constant goes to infinity and large $N_c$ limit. However,
with a finite value of $\lambda$ and the $N_c=3$, an understanding of how these computations are affected by finite $\lambda$
corrections may be essential for more precise theoretical predictions.
There are two sources of contributions to finite $\lambda$ corrections: 1) The string theory corrections of the
AdS-Schwarzschild background where the world sheet underlying the jet quenching Wilson loop is embedded. 2) The fluctuation
of the world sheet itself. The former has been calculated in \cite{Armesto} and it amounts to multiply $\hat q_{SYM}^{(0)}$
by a factor
$1-1.765\lambda^{-3/2}$. The parallel correction within the
framework of the drag force has also been figured out
in \cite{JF}, and has been found to enhance the jet quenching parameter. The latter contribution, that from the fluctuations, however, is of the order $O(\lambda^{-1/2})$ and thereby dominates the former. But it is much harder to evaluate. It is the purpose of the present paper to determine this contribution and we obtain the corrected jet-quenching parameter
\begin{equation}
\hat q_{SYM}(\lambda)=\hat q_{SYM}^{(0)}[1+\kappa\lambda^{-1/2}+O(\lambda^{-1})]
\end{equation}
with the value of $\kappa$
\begin{equation}
\kappa\simeq-1.97.
\label{result}
\end{equation}

The paper is organized as follows. In the next section, we will derive the partition function with the world fluctuations.
We will present our analytical study and numerical results in section 3 and section 4 respectively. The section 5 concludes
the paper along with some discussions of the result and some open issues. Some calculation technique details are presented
in the appendixes.

\section{The one loop effective action}

It follows from the AdS/CFT correspondence that the expectation value of a Wilson loop of the $\mathcal N=4$ SYM
in the fundamental representation of the gauge group is related to the path integral of a type IIB superstring
in a 10-dimensional spacetime that is asymptotically $AdS_5\times S^5$, i. e.
\begin{equation}
<W[{\cal C}]> = {const.}\int[dX][d\theta]e^{iS[X,\theta]}\equiv e^{iS_{\rm eff.}[{\cal C}]},
\label{def}
\end{equation}
where $X$'s and $\theta$'s are the bosonic coordinates and fermionic coordinates, and the superstring action
\begin{equation}
S[X,\theta]=S_B[X]+S_F[\theta]
\label{superstring}
\end{equation}
with $S_B[X]$ and $S_F[\theta]$ the bosonic and fermionic parts \cite{Metsaev, Cvetic}.
The semiclassical approximation of the effective action $S_{\rm eff.}$ reads
\begin{equation}
S_{\rm eff.}[{\cal C}]=\sqrt{\lambda}\Big[S_0[{\cal C}]
+\frac{S_1[{\cal C}]}{\sqrt{\lambda}}+O\left(\frac{1}{\lambda}\right)\Big],
\end{equation}
where $\sqrt{\lambda}S_0[{\cal C}]\equiv S[\bar X,0]$ with $\bar X$ the classical solution and $S_1[{\cal C}]$ stands for the one-loop correction. With the light-like antiparallel lines,
the jet-quenching parameter comes from the second power of the Taylor expansion
of $S_{\rm eff.}$ in
\begin{equation}
\epsilon=\frac{(2\pi)^{\frac{3}{2}}TL}{\Gamma^2\left(\frac{1}{4}\right)}<<1
\end{equation}
with $L$ the distance between the two lines and $T$ the temperature.

To figure out $S_1[{\cal C}]$, we need to expand $S[X,\theta]$ to the quadratic order around the classical solution
$\bar X$ and $\theta=0$. To the same order of $\lambda$, the 10-dimensional metric
takes the form of Schwarzschild-$AdS_5\times S^5$
\begin{eqnarray}
ds^2&=&-r^2(1+f)dx^+dx^-+\frac{r^2}{2}(1-f)[(dx^+)^2+(dx^-)^2]+r^2[(dx^1)^2+(dx^2)^2]
+\frac{dr^2}{r^2f}+d\Omega_5^2\nonumber\\
&=&G_{\mu\nu}(X)dX^\mu dX^\nu,
\label{target}
\end{eqnarray}
where $f=1-\frac{r_h^4}{r^4}$, $d\Omega_5$ is the line element on $S^5$ and $x^\pm$
corresponds to the light-cone coordinates on the boundary ($r\to\infty$). In the
last step of (\ref{target}), we have introduced the compact notation of the
target spacetime coordinates
\begin{equation}
X^\mu=(x^+,x^-,x^1,x^2,r;\Omega_5)
\end{equation}
with $\Omega_5$ the set of five coordinates on $S^5$.

The bosonic part of (\ref{superstring}) $S_B[X]$ is the Nambu-Goto action of a 2d world sheet embedded in the
target spacetime, given by
\begin{equation}
S_B[X]=\frac{1}{2\pi\alpha'}\int d\tau\int dr\sqrt{-g},
\label{boson}
\end{equation}
where $g$ is the determinant of the induced 2d metric
\begin{equation}
g_{\alpha\beta}=G_{\mu\nu}(X)\frac{\partial X^\mu}{\partial\sigma^\alpha}\frac{\partial X^\nu}{\partial\sigma^\beta},
\end{equation}
with $\sigma^\alpha(\alpha=0,1)$ the world sheet parametrization.
The fermionic part of (\ref{superstring}), $S_F[\theta]$, takes the form \cite{Metsaev, Cvetic, Townsend}
\begin{equation}
S_F[\theta]=\frac{1}{2\pi\alpha^{'}}\int d^2\sigma(2\sqrt{-g}g^{\alpha\beta}\bar{\theta}
\rho_\alpha D_\beta\theta-i\epsilon^{\alpha\beta}\bar{\theta}\rho_\alpha\rho_\beta \theta)
\label{fermion}
\end{equation}
where $\theta$ and $\bar\theta$ are two $16\times 1$ Majorana-Weyl spinors in the target spacetime and
\begin{equation}
\rho_\alpha\equiv\eta_\alpha^\mu E_\mu^a\Gamma_a \qquad \eta_\alpha^\mu=\frac{\partial X^\mu}{\partial \sigma^\alpha}
\end{equation}
with $\Gamma_a(a=0,1,2,3,4)$ the $16\times 16$ Gamma matrices satisfying
\begin{equation}
\{{\Gamma^a,\Gamma^b}\}=2\eta^{ab}
\end{equation}
and $\eta={\rm diag}(-1,1,1,1,1)$. The higher powers of $\theta$ has been ignored and the bosonic coordinates
in $S_F[\theta]$ has been approximated by the classical solution. The Majorana condition relates $\theta$ and
$\bar\theta$ via
\begin{equation}
\bar\theta=\tilde\theta C\times C
\end{equation}
with $C$ the $4\times 4$ charge conjugation matrix.
The $\kappa$-symmetry of the original fermionic action up the the order $\theta^2$ in the Schwarzschild
$AdS_5\times S^5$ background has been employed to reduce the fermionic degrees of freedom by half.
We refer the interested readers to \cite{HJR} for the detailed reduction for the present application.
It is a straightforward generalization of the $\kappa$-symmetry in the $AdS_5\times S^5$ background \cite{IP,RK,RK1}.

The classical world sheet corresponds to
\begin{equation}
\bar X^\mu=(0,\tau,0,x^2,r,0,0,0,0,0)
\label{classical}
\end{equation}
and $\theta=0$ with the function $x^2$ given implicitly by
\begin{equation}
\frac{dx^2}{dr}=\pm\frac{\epsilon}{r^2\sqrt{f}},
\end{equation}
where
\begin{equation}
\epsilon=\frac{(2\pi)^{\frac{3}{2}}TL}{\Gamma^2\left(\frac{1}{4}\right)}.
\end{equation}
\begin{figure}[h]
\centering
\includegraphics[height=2.0in]{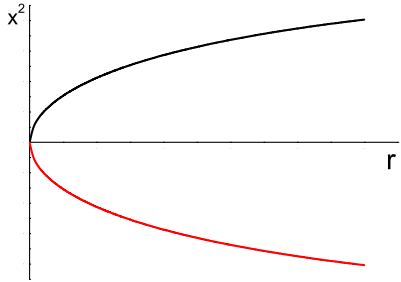}
\caption{The curve represents $x^2$ vs r.}
\end{figure}
and the positive(negative) sign refers to the upper(lower) branch in FIG.1. The projection of
the classical word sheet onto $S^5$ is a point.
The induced world sheet metric
\begin{equation}
ds^2=\bar g_{\alpha\beta}d\sigma^\alpha d\sigma^\beta=\frac{r_h^4}{2r^2}d\tau^2+\frac{1+\epsilon^2}{r^2f}dr^2
\label{spacelike}
\end{equation}
with $\sigma=(\tau,r)$, and
\begin{equation}
\bar g_{\alpha\beta}=G_{\mu\nu}\frac{\partial\bar X^\mu}{\partial\sigma^\alpha}\frac{\partial\bar X^\nu}{\partial\sigma^\beta}.
\end{equation}
and the world sheet scalar curvature reads
\begin{equation}
R=-\frac{2}{1+\epsilon^2}(1-\frac{3r_h^4}{r^4}),
\end{equation}
and one can readily verify that
\begin{equation}
\int d^2\sigma\sqrt{\bar{g}}R=0.
\label{interR}
\end{equation}
The classical action reads
\begin{equation}
S[\bar X,0]=S_B[\bar X]=i\frac{\Gamma^2\left(\frac{1}{4}\right)}{8\sqrt{\pi}}\sqrt{\lambda}L_-T\sqrt{1+\epsilon^2}
\end{equation}
whose $\epsilon^2$ term of the Taylor expansion gives rise to (\ref{leading}) following the definition (\ref{jet})\\
($<W^A[{\cal C}]>\simeq <W[{\cal C}]>^2$ at large $N_c$). Because the metric of the classical world sheet
(\ref{spacelike}) is {\it space-like}, its Nambu-Goto action and the quadratic action
of the fluctuations around it are all {\it imaginary}.

For the rest part of this section, it is convenient to introduce the funfbeins pertaining tangent space of the
Schwarzschild-$AdS_5$ sector
\begin{eqnarray}
E^0&=&r\sqrt{\frac{f}{2}}(dX^++dX^-)\nonumber\\
E^3&=&\frac{r}{\sqrt{2}}(dX^+-dX^-)\nonumber\\
E^i&=&rdX^i,i=1,2\nonumber\\
E^4&=&\frac{1}{r\sqrt{f}}dX^4,
\end{eqnarray}
which can be further decomposed into transverse and longitudinal components via a rotation
\begin{eqnarray}
{\cal E}^0 &=& E^0\cosh\beta+E^3\sinh\beta\nonumber\\
{\cal E}^3 &=& E^0\sinh\beta+E^3\cosh\beta\nonumber\\
{\cal E}^1 &=& E^1\nonumber\\
{\cal E}^2 &=& E^2\cos\alpha-E^4\sin\alpha\nonumber\\
{\cal E}^4 &=& E^2\sin\alpha+E^4\cos\alpha
\end{eqnarray}
with $\tanh\beta=\sqrt{f}$ and $\tan\alpha=\epsilon$.

The action underlying the partition function consists of the the quadratic terms of $S[\bar X+\delta X,\theta]$
in $\delta X$ and $\theta$  with
\begin{equation}
\delta X^\mu(\sigma)=(\delta X^+,\delta X^-,\delta X^1,\delta X^2,\delta X^4,\delta\Omega_5).
\end{equation}
The linear terms vanish because of the classical equation of motion.

\subsection{Bosonic fluctuations}

Among five fluctuation coordinates within Schwarzschild-$AdS_5$ sector, only three of them are physical and the
rest two correspond to a world sheet diffeomophism. This can be made explicit be projecting
$\delta X$ onto the funfbeins ${\cal E}$'s, i.e. $z^a\equiv E_\mu^a\delta X^\mu(a=0,1,2,3,4)$ and
to retain only the transverse fluctuations $z^0$ and $z^1$ and $z^2$. As the longitudinal fluctuations
$z^3$ and $z^4$ contribute only zero
modes of the quadratic action, it is more convenient to work within the static gauge,
$\delta X^-=\delta X^4=0$, through a superposition of the transverse and longitudinal ones. Then $G_{\mu\nu}$ is {\it independent} of $\delta X$ and we have
\begin{equation}
g_{\alpha\beta}=\bar g_{\alpha\beta}
+G_{\mu\nu}(\bar X)\frac{\partial\delta X^\mu}{\partial\sigma^\alpha}\frac{\partial\delta X^\nu}{\partial\sigma^\beta}.
\end{equation}
Upon the parametrization
\begin{equation}
u\equiv\delta X^+=\frac{1}{r}\sqrt{\frac{1-f}{2f}}\xi,v\equiv\delta X^2=\frac{\sqrt{1+\epsilon^2}}{r}\eta,
w\equiv\delta X^1=\frac{1}{r}\zeta
\end{equation}
and $\delta\Omega_5={\delta_s}$ with $s=1,...,5$ with $\chi_s$ the five tangent coordinates
in the neighborhood of the classical point. Expanding the bosonic action (\ref{boson})
with $\sqrt{-g}=\sqrt{-g_{\tau\tau}g_{rr}+g_{\tau r}^2}$
to the quadratic order in $\delta X^\mu$, we find that
\begin{equation}
S_{bosonic}=\frac{1}{2\pi\alpha'}\left(
\hat S_\zeta+\hat S_{\xi\eta}+\hat S_\delta\right)
\end{equation}
where
\begin{equation}
\hat S_\zeta=\frac{1}{2}i\int d\tau dr\sqrt{\bar{g}}{(\bar{g} ^{\tau\tau}\dot{\zeta} ^2+
\bar{g} ^{rr}{\zeta^{'}}^2+M_\zeta^2\zeta^2})\equiv i\tilde\zeta{\cal A}_\zeta\zeta
\label{zeta}
\end{equation}
\begin{eqnarray}
\hat S_{\xi\eta}&=&\frac{1}{2}i\int d\tau dr\sqrt{\bar{g}}(-\bar{g} ^{\tau\tau}\dot{\xi} ^2-
\bar{g} ^{rr}{\xi^{'}}^2+\bar{g} ^{\tau\tau}\dot{\eta} ^2+
\bar{g} ^{rr}{\eta^{'}}^2-M_\xi^2\xi^2+M_\eta^2\eta^2-\frac{8\sqrt{2}\epsilon}{\sqrt{1+\epsilon^2}}\frac{r}{r_h^2}\xi\dot{\eta})
\nonumber\\
&\equiv&\frac{1}{2}i(\tilde\xi,\tilde\eta){\cal A}_{\xi\eta}\left(\begin{array}{cc} \xi \\ \eta \end{array}\right)
\label{xieta}
\end{eqnarray}
and
\begin{equation}
\hat S_\delta=\frac{1}{2}i\int d\tau dr\sqrt{\bar g}
\left(\bar{g}^{\tau\tau}\dot{\delta_s}^2+
\bar{g}^{rr}{\delta_s'}^2\right)
\label{u}
\end{equation}
where
$\bar{g}^{\tau\tau}=\frac{2r^2}{r_h^4},\bar{g} ^{rr}=\frac{r^4-r_h^4}{r^2(1+\epsilon^2)}$, $\dot{F}\equiv\frac{\partial F}{\partial\tau}$ and $F'\equiv\frac{\partial F}{\partial r}$. The
diagonal masses in (\ref{zeta}) and (\ref{xieta}) read
\begin{equation}
M_\zeta^2=M_\eta^2=\frac{\bar{g} ^{rr}}{r^2}+\frac{(\sqrt{\bar{g}}\bar{g} ^{rr}/r)^{'}}{\sqrt{\bar{g}}}=\frac{2r_h^4}{r^4(1+\epsilon^2)}=\frac{M_1^2}{1+\epsilon^2}
\end{equation}
and
\begin{equation}
M_\xi^2=\frac{2}{1+\epsilon^2}(3+\frac{r_h^4}{r^4})=\frac{M_2^2}{1+\epsilon^2}.
\end{equation}
Notice the negative sign in front of the terms quadratic in $\xi$.

\subsection{Fermionic fluctuations}

To simplify the fermionic action (\ref{fermion}), we take the representation
\begin{equation}
\Gamma^a=\gamma^a\times I_4
\end{equation}
with $\gamma^a$ $4\times 4$ gamma matrices satisfying
\begin{equation}
\{{\gamma^a,\gamma^b}\}=2\eta^{ab}
\end{equation}

Upon a similarity transformation $\theta=V\psi$ with
\begin{equation}
V=e^{-\frac{1}{2}\beta\Gamma_0\Gamma_3+\frac{1}{2}\alpha\Gamma_2\Gamma_4}
\end{equation}
we have
\begin{equation}
V^{-1}(\bar{g}^{ij}\rho_iD_j)V=\Gamma^\tau(\partial_\tau-\frac{1}{2}\omega_\tau^{01}
\Gamma_3\Gamma_4)+\Gamma^r\partial_r+\frac{\epsilon}{\sqrt{1+\epsilon^2}}\Gamma_0\Gamma_2
\Gamma_3
\end{equation}
where
\begin{equation}
\Gamma^\tau=-e^{0\tau}\Gamma_3 \qquad \Gamma^r=e^{1r}\Gamma_4
\end{equation}
with the world sheet zweibines and spin connections given by
\begin{equation}
e_\tau^0=\frac{r_h^2}{\sqrt{2}r} \qquad e_r^1=\frac{1}{r}\sqrt{\frac{1+\epsilon^2}{f}}
\end{equation}
and
\begin{equation}
\omega_\tau^{01}=-\omega_\tau^{10}=-\frac{r_h^2}{r}\sqrt{\frac{f}{2(1+\epsilon^2)}}.
\end{equation}
The fermionic action in terms of $\psi$ reads then
\begin{equation}
S_F[\theta]=\frac{i}{\pi\alpha^{'}}\int d^2\sigma\sqrt{\bar{g}}\bar{\psi}[\Gamma^\tau(
\partial_\tau-\frac{1}{2}\omega_\tau^{01}\Gamma_3\Gamma_4)+\Gamma^r\partial_r+
i\Gamma_3\Gamma_4+\frac{\epsilon}{\sqrt{1+\epsilon^2}}\Gamma_0\Gamma_2
\Gamma_3]\psi.
\end{equation}

The next step is to decompose the $16\times 1$ $\psi$ into eight world sheet
Majorana fermions. This is accomplished with the representation
\begin{eqnarray}
\gamma_0&=&\left(\begin{array}{cc} 0 & i\sigma_2 \\ i\sigma_2 & 0 \end{array}\right)
\qquad
\gamma_1=\left(\begin{array}{cc} \sigma_2 & 0 \\ 0 & -\sigma_2 \end{array}\right)\nonumber\\
\gamma_2&=&\left(\begin{array}{cc} 0 & -i\sigma_2 \\ i\sigma_2 & 0 \end{array}\right)
\qquad
\gamma_3=\left(\begin{array}{cc} \sigma_3 & 0 \\ 0 & \sigma_3 \end{array}\right)\nonumber\\
\gamma_4&=&\left(\begin{array}{cc} -\sigma_1 & 0 \\ 0 & -\sigma_1 \end{array}\right)
\end{eqnarray}
with $\sigma_i(i=1,2,3)$ Pauli matrices. It follows that
\begin{eqnarray}
&&\Gamma_3=\left(\begin{array}{cc} \sigma_3\times I_4 & 0 \\ 0 & \sigma_3\times I_4 \end{array}\right)
\qquad
\Gamma_4=\left(\begin{array}{cc} -\sigma_1\times I_4 & 0 \\ 0 & -\sigma_1\times I_4 \end{array}\right)\nonumber\\
&&\Gamma_3\Gamma_4=\left(\begin{array}{cc} -i\sigma_2\times I_4 & 0 \\ 0 & -i\sigma_2\times I_4 \end{array}\right)
\qquad
\Gamma_0\Gamma_2\Gamma_3=\left(\begin{array}{cc} -\sigma_3\times I_4 & 0 \\ 0 & \sigma_3\times I_4 \end{array}\right)
\end{eqnarray}
and
\begin{equation}
S_F[\chi]=\frac{i}{\pi\alpha'}\sum_{s=1}^4\int d\tau dr\sqrt{\bar g}\left(\bar\chi_{+s}{\cal A}_+\chi_{+s}
+\bar\chi_{-s}{\cal A}_-\chi_{-s}\right),
\label{wfermion}
\end{equation}
where
\begin{eqnarray}
{\cal A}_\pm=-e^{0\tau}\sigma_3\left(\frac{\partial}{\partial\tau}+\frac{i}{2}\omega_\tau^{01}\right)
-e^{1r}\sigma_1\frac{\partial}{\partial r}+i\sigma_2\mp\frac{\epsilon}{\sqrt{1+\epsilon^2}}\sigma_3.
\end{eqnarray}
and $\chi_{\pm s}$'s are two component Majorana fermions.

In summary, the explicit form of the fluctuation action reads
\begin{equation}
\Delta S[\xi,\eta,\zeta,\delta,\chi]\equiv\frac{1}{2\pi\alpha'}(\hat S_{\xi\eta}+\hat S_\zeta+\hat S_\delta)+iS_F[\chi]
\end{equation}
with $\hat S_\zeta$, $\hat S_{\xi\eta}$, $\hat S_\delta$ and $S_F[\theta]$ given by (\ref{zeta}), (\ref{xieta}), (\ref{u}) and
(\ref{wfermion}).
The partition function of the fluctuations is defined to be the path integral of
\begin{equation}
{\cal Z}={\rm const.}\int [d\xi][d\eta][d\zeta][d\delta][d\chi]e^{i\Delta S[\xi,\eta,\zeta,\delta,\chi]}
\label{partition1}
\end{equation}
and the effective action up to one-loop order reads
\begin{equation}
S_{\rm eff.}=S[\bar X,0]-i\ln {\cal Z}.
\label{eff1}
\end{equation}

As was pointed out in \cite{Philip}, the classical world sheet underlying (\ref{leading}) is the
saddle point of the Nambu-Goto action. Therefore the bosonic part of the path integral (\ref{partition1}) is
problematic. In particular, the path integral
\begin{equation}
\int [d\xi][d\eta]e^{i\frac{\hat S_{\xi\eta}}{2\pi\alpha'}}
=\int [d\xi][d\eta]e^{-\frac{1}{4\pi\alpha'}
(\tilde\xi,\tilde\eta){\cal A}_{\xi\eta}\left(\begin{array}{cc} \xi \\ \eta \end{array}\right)}
\label{pathxieta}
\end{equation}
with ${\cal A}_{\xi\eta}$ defined in (\ref{xieta}) does not exist for a real and positive $\alpha'$
because of the negative eigenvalues of ${\cal A}_{\xi\eta}$.
In order to make the leading order term (\ref{leading}) stable and to make the one-loop correction to be computed
below meaningful, we {\it define} the path integral (\ref{partition1}) for $\alpha'>0$
as an analytical continuation from an imaginary $\alpha'$.
In terms of the spectrum of ${\cal A}_{\xi\eta}$, $\{{\omega_n}\}$, and $\alpha'=is$ with $s>0$, the path integral
(\ref{pathxieta}) becomes
\begin{equation}
\int [d\xi][d\eta]e^{i\frac{\hat S_{\xi\eta}}{2\pi\alpha'}}
=\prod_n\left(2\pi e^{\pm i\frac{\pi}{4}}\sqrt{\frac{s}{|\omega_n|}}\right)
\propto\left|{\rm det}^{-\frac{1}{2}}{\cal A}_{\xi\eta}\right|
\end{equation}
with the plus(minus) sign corresponding to a positive(negative) $\omega_n$.
Though we are not able to come up with a physical interpretation of this
procedure, we observe that the sign ambiguity associated to the continuation of the square root
$\sqrt{s}=\sqrt{-i\alpha'}$ from the imaginary $\alpha'$ back to a real and positive one
may be removed by renormalizing (\ref{partition1}) by its value at $\epsilon=0$. This amounts to subtracting
from (\ref{eff1}) the one-loop effective action of two isolated light-like straight lines. Since
the jet quenching parameter comes from the $\epsilon^2$ term of the Taylor expansion of (\ref{eff1}), this
renormalization is unnecessary for our purpose.
Choosing the constant in front of the path integral of (\ref{partition1}) appropriately, we end up with
\begin{equation}
{\cal Z}=\left|\frac{{\rm det}^2{\cal A}_+{\rm det}^2{\cal A}_-}{{\rm det}^{\frac{1}{2}}
{\cal A}_{\xi\eta}{\rm det}^{\frac{1}{2}}{\cal A}_{\zeta}{\rm det}^{\frac{5}{2}}(-\nabla^2)}\right|.
\label{partition}
\end{equation}
and the finite 't Hooft coupling correction to the jet quenching parameter comes from the $\epsilon^2$ term of
the Taylor expansion of
\begin{equation}
z\equiv-\lim_{{L_-}\to\infty}\frac{\ln{\cal Z}}{{L_-}},
\end{equation}
where ${L_-}$ defines the domain of $\tau$, i.e. $0<\tau<{L_-}$.

Because of the cancelation between bosonic and fermionic degrees of freedom, $S_{\rm eff}$ is at most
logarithmically divergent. The coefficient of the divergence takes the invariant form
\begin{equation}
\int d^2\sigma\sqrt{\bar g}(c_1+c_2R)
\label{uvmass}
\end{equation}
It is shown in the appendix A that $c_1=0$ as expected. While $c_2\neq 0$ in the static gauge, this is not
bothering because of (\ref{interR}) for the world sheet under consideration. We will not address the subtleties brought
about by the world sheet boundary \cite{orlando} but are content with an explicit demonstration that the one loop correction to the jet quenching coefficient is indeed divergence free in the next section.

\section{The formula for $\kappa$}

In this section, we shall transform the effective action (\ref{eff1}) into the form suitable for the numerical
calculations described in the subsequent section. In the course of the reduction, we shall demonstrate explicitly
that the correction to the jet quenching parameter is free from both ultraviolet and infrared divergence.

It is convenient to work with the new coordinates
\begin{equation}
t=\frac{r_h\tau}{\sqrt{1+\epsilon^2}} \qquad \rho=\pm\frac{\sqrt{2}}{r_h}\int_{r_h}^r\frac{dr'}{\sqrt{1-r_h^4/r^{'4}}}
\end{equation}
where the section $x_2>0$(the upper branch of FIG.1) of the world sheet is mapped to $0<\rho<\infty$ and the section
$x_2<0$(the lower branch of FIG.1) to
$-\infty<\rho<0$. We have
\begin{equation}
\rho\simeq\pm\frac{\sqrt{2}r}{r_h}
\label{rho}
\end{equation}
for large $r$ and
\begin{equation}
\rho\simeq\pm\sqrt{2}\frac{\sqrt{r-r_h}}{r_h^{\frac{1}{2}}}.
\end{equation}
for $r$ near the horizon. So the two branches of $r$ to $\rho$ mapping join smoothly at $\rho=0$. In terms of
the coordinates ($t$, $\rho$), the world sheet metric takes the conformal form
\begin{equation}
ds^2=(1+\epsilon^2)e^{2\phi}(dt^2+d\rho^2)
\end{equation}
with
\begin{equation}
e^{2\phi}=\frac{r_h^2}{2r^2}.
\end{equation}
The corresponding zweibines and spin connections read
\begin{equation}
e^0=\sqrt{1+\epsilon^2}e^\phi dt \qquad e^1=\sqrt{1+\epsilon^2}e^\phi d\rho
\end{equation}
and
\begin{equation}
\omega_t^{01}=-\frac{r_h^2}{r}\sqrt{\frac{f}{2}}dt=\frac{d\phi}{d\rho}dt.
\end{equation}
The functional operators underlying the determinants in (\ref{eff1}) becomes
\begin{eqnarray}
{\cal A}_\zeta=\frac{\hat{{\cal A}}_\zeta}{1+\epsilon^2}\qquad
{\cal A}_{\xi\eta}=\frac{\hat{{\cal A}}_{\xi\eta}}{1+\epsilon^2}\qquad
{\cal A}_u=\frac{\hat{{\cal A}}_u}{1+\epsilon^2}\qquad
{\cal A}_\pm=\frac{\hat{{\cal A}}_\pm}{\sqrt{1+\epsilon^2}},
\end{eqnarray}
where
\begin{eqnarray}
\hat{{\cal A}}_\zeta &=& -e^{-2\phi}\left(\frac{\partial^2}{\partial t^2}+\frac{\partial^2}{\partial \rho^2}\right)
                     +M_1^2\nonumber\\
\hat{{\cal A}}_{\xi\eta} &=&
\left(\begin{array}{cc}
     e^{-2\phi}\left(\frac{\partial^2}{\partial t^2}+\frac{\partial^2}{\partial \rho^2}\right)+M_2^2 &
     4\epsilon e^{-\phi}\frac{\partial}{\partial t} \\
     -4\epsilon e^{-\phi}\frac{\partial}{\partial t} &
     -e^{-2\phi}\left(\frac{\partial^2}{\partial t^2}+\frac{\partial^2}{\partial \rho^2}\right)+M_1^2
\end{array}\right)\nonumber\\
\hat{{\cal A}}_u &=& -e^{-2\phi}\left(\frac{\partial^2}{\partial t^2}+\frac{\partial^2}{\partial \rho^2}\right)\nonumber\\
\hat{{\cal A}}_\pm &=& -e^{-\phi}\Big[\sigma_3\left(\frac{\partial}{\partial t}+\frac{i}{2}\frac{d\phi}{d\rho}\sigma_2\right)
                                  +\sigma_1\frac{\partial}{\partial\rho}\Big]
                   +i\sigma_2\sqrt{1+\epsilon^2}\mp\epsilon\sigma_3.
\label{hatop}
\end{eqnarray}
The partition function becomes then
\begin{equation}
{\cal Z}=\left|\frac{{\rm det}^2\hat{{\cal A}}_+{\rm det}^2\hat{{\cal A}}_-}
{{\rm det}^{\frac{1}{2}}\hat{{\cal A}}_{\xi\eta}{\rm det}^{\frac{1}{2}}\hat{{\cal A}}_{\zeta}
{\rm det}^{\frac{5}{2}}\hat{\cal A}_u}\right|.
\label{hateff}
\end{equation}
Upon a Fourier transformation in $t$, we find that
\begin{equation}
{\cal Z}=\prod_{\omega}Z(\omega)
\end{equation}
with
\begin{equation}
Z(\omega)=\left|\frac{{\rm det}^2\hat{A}_+(\omega){\rm det}^2\hat{A}_-(\omega)}
{{\rm det}^{\frac{1}{2}}\hat{{\cal A}}_{\xi\eta}(\omega){\rm det}^{\frac{1}{2}}\hat{{\cal A}}_{\zeta}(\omega)
{\rm det}^{\frac{5}{2}}\hat{{\cal A}}_{u}(\omega)}\right|,
\label{hateff}
\end{equation}
and $z=\frac{1}{{L_-}}\sum_{\omega}\ln Z(\omega)$,
where $A_{...}(\omega)$ is obtained from ${\cal A}_{...}$ of (\ref{hatop}) by replacing $\frac{\partial}{\partial t}$
with $-i\omega$. Explicitly, we have
\begin{eqnarray}
\hat{A}_\zeta(\omega) &=& e^{-2\phi}D_1(\omega)\nonumber\\
\hat{A}_{\xi\eta}(\omega) &=& e^{-2\phi}D_{\xi\eta}(\omega)\nonumber\\
\hat{{\cal A}}_u(\omega)&=&-e^{-2\phi}D_0(\omega)\nonumber\\
\hat{{\cal A}}_\pm(\omega)&=&-e^{-\frac{3}{2}\phi}D_\pm(\omega)e^{\frac{1}{2}\phi},
\label{scaling}
\end{eqnarray}
where the functional operators
\begin{eqnarray}
D_0(\omega)&=&-\frac{d^2}{d\rho^2}+\omega^2\nonumber\\
D_1(\omega)&=&-\frac{d^2}{d\rho^2}+\omega^2+e^{2\phi}M_1^2\nonumber\\
D_{\xi\eta}(\omega) &=& \left(\begin{array}{cc} -D_2(\omega) & 4i\epsilon\omega e^\phi\\
                                     -4i\epsilon\omega e^\phi & D_1(\omega) \end{array}\right)\nonumber\\
D_\pm(\omega) &=& -\sigma_1\frac{d}{d\rho}+i\omega\sigma_3+e^\phi(i\sqrt{1+\epsilon^2}\sigma_2\mp\epsilon\sigma_3)
\label{det}
\end{eqnarray}
with
\begin{eqnarray}
D_2(\omega)=-\frac{d^2}{d\rho^2}+\omega^2+e^{2\phi}M_2^2
\label{det3}
\end{eqnarray}
Taking the square of $D_\pm(\omega)$, we find
\begin{eqnarray}
D_\pm^2(\omega)&=& \frac{d^2}{d\rho^2}-\omega^2\mp2i\epsilon\omega e^\phi-e^{2\phi}
		      +(\sqrt{1+\epsilon^2}\sigma_3\mp i\epsilon\sigma_2)e^\phi\frac{d\phi}{d\rho}
                \nonumber\\&=& -V_\pm{\cal D}_\pm(\omega)V_\pm^{-1}
\end{eqnarray}
where $V_\pm$ is the $2\times 2$ matrix that diagonalizes the matrix $\sqrt{1+\epsilon^2}\sigma_3\mp i\epsilon\sigma_2$ and
\begin{eqnarray}
{\cal D}_\pm(\omega) &=& -\frac{d^2}{d\rho^2}+\omega^2\pm 2i\epsilon\omega e^\phi+e^{2\phi}
-\sigma_3 e^\phi\frac{d\phi}{d\rho}\nonumber\\
&=& \left(\begin{array}{cc} D_3(\omega)\mp 2i\epsilon\omega e^\phi & 0\\
                            0 & D_3^\prime(\omega)\mp 2i\epsilon\omega e^\phi \end{array}\right)
\label{det2}
\end{eqnarray}
with
\begin{eqnarray}
D_3(\omega) &=& -\frac{d^2}{d\rho^2}+(\omega^2+e^{2\phi}-e^\phi\frac{d\phi}{d\rho})\nonumber\\
D_3^\prime(\omega) &=& -\frac{d^2}{d\rho^2}+(\omega^2+e^{2\phi}+e^\phi\frac{d\phi}{d\rho})
\label{det1}
\end{eqnarray}
Replacing the $\rho$ in $D_3(\omega)$ by $-\rho$, we can get $D_3^\prime(\omega)$ and therefore ${\rm det}D_3(\omega)={\rm det}D_3^\prime(\omega)$.

The scaling factors $e^{-2\phi}$, $e^{-\phi}$, $e^{-\frac{3}{2}\phi}$ and $e^{\frac{1}{2}\phi}$ of (\ref{scaling})
cancel in the ratio of determinants of (\ref{hateff}). Consequently
\begin{equation}
Z(\omega)=\left|\frac{\rm det {\cal D}_+(\omega)\rm det {\cal D}_-(\omega)}
{{\rm det}^{\frac{1}{2}}D_{\xi\eta}(\omega){\rm det}^{\frac{1}{2}}D_1(\omega)
{\rm det}^{\frac{5}{2}}D_0(\omega)}\right|.
\label{z}
\end{equation}

In the limit ${L_-}\to\infty$,
\begin{equation}
\sum_{\omega}(...)\to {L_-}^\prime\int_{-\infty}^\infty\frac{d\omega}{2\pi}(...)
\end{equation}
with ${L_-}^\prime=\frac{r_h}{\sqrt{1+\epsilon^2}}{L_-}$ the length of the domain of $t$. It follows that
\begin{equation}
z=-\frac{T}{2\sqrt{1+\epsilon^2}}\int_{-\infty}^\infty d\omega\ln Z(\omega).
\end{equation}
Since $Z(\omega)$ is an even function of $\epsilon$, we have the Taylor expansion
\begin{equation}
\ln Z(\omega)=a(\omega)+b(\omega)\epsilon^2+O(\epsilon^4)
\label{exp}
\end{equation}
and
\begin{equation}
z=-\frac{T}{2}\int_{-\infty}^\infty d\omega a(\omega)
+\epsilon^2\frac{T}{2}\int_{-\infty}^\infty d\omega
\Big[\frac{1}{2}a(\omega)-b(\omega)\Big]+O(\epsilon^4).
\end{equation}
It is the second integral that contributes to the correction to the jet quenching parameter.

Hence, the strong coupling expansion of the jet quenching parameter can be written as
\begin{equation}
\hat q_{SYM}=\hat q_{SYM}^{(0)}[1+\kappa\lambda^{-1/2}+O(\lambda^{-1})]
\end{equation}
where
\begin{equation}
\kappa=\frac{8\sqrt{\pi}}{\Gamma^2{(\frac{1}{4})}}\int_{-\infty}^\infty
d\omega\Big[\frac{1}{2}a(\omega)-b(\omega)\Big].
\label{k}
\end{equation}

\subsection{The formulae of $a(\omega)$}

The function $a(\omega)$ corresponds to $\ln Z(\omega)$ at $\epsilon=0$ and we have
\begin{equation}
a(\omega)=\ln\left|\frac{{\rm det}^4D_3(\omega)}{{\rm det}D_1(\omega){\rm det}^{\frac{1}{2}}D_2(\omega)
{\rm det}^{\frac{5}{2}}D_0(\omega)}\right|.
\label{aratio}
\end{equation}
$a(\omega)$ can be factorized as a product of determinant ratios, which can be evaluated by the Gelfand-Yaglom method
\cite{Gelfand}\cite{Drukker:2000ep}. We give a brief introduction here. The examples of its application to other Wilson
loops can be found in \cite{kruczenski, chu:2009, foroni, zhang2011}.

Consider two functional operators
\begin{equation}
H_n = -\frac{d^2}{dx^2}+V_n(x),
\end{equation}
with $n=1,2$, defined in the domain $a\le x\le b$ under the Dirichlet boundary condition, the determinant ratio
\begin{equation}
\frac{{\rm det}H_2}{{\rm det}H_1}=\frac{f_2(b|a)}{f_1(b|a)}
\label{dratio}
\end{equation}
where $f_n(x|a)$ is the solution of the homogeneous equation
\begin{equation}
H_n f_n=0,
\label{homo}
\end{equation}
subject to the conditions $f_n(0|a)=0$ and
$f_n^\prime(0|a)=1$. In terms of a pair of linearly independent
solutions of (\ref{homo}), $(\eta_n,\xi_n)$
\begin{equation}
f_n(b|a)=\frac{\eta_n(a)\xi_n(b)-\eta_n(b)\xi_n(a)}
{W[\eta_n,\xi_n]}.
\label{lambda}
\end{equation}
where the Wronskian $W[\eta_n,\xi_n]$ is $x$-independent.

For our problem, we have $H_1=D_0(\omega)$, $D_1(\omega)$ or
$D_2(\omega)$ and $H_2=D_3(\omega)$ under the limit $a\to-\infty$ and $b\to\infty$.
Let us define by $u_\alpha^{(1)}$ and $u_\alpha^{(2)}$ the two linearly independent solutions of the equations
$D_\alpha\psi=0$($\alpha=0,1,2,3$) subject to the boundary conditions
\begin{eqnarray}
u_\alpha^{(1)}(\rho)=\{\begin{array}{cc}e^{\omega\rho}  & \rho\rightarrow-\infty \\ C_\alpha(\omega)e^{\omega\rho} & \rho\rightarrow\infty\end{array}\nonumber\\
u_\alpha^{(2)}(\rho)=\{\begin{array}{cc}C_\alpha(\omega)e^{-\omega\rho}  & \rho\rightarrow-\infty \\  e^{-\omega\rho} & \rho\rightarrow\infty\end{array}
\label{solution}
\end{eqnarray}
We have $C_0(\omega)=1$ and the rest of $C_\alpha(\omega)$'s satisfy
\begin{equation}
W[u_\alpha^{(1)},u_\alpha^{(2)}]=-2\omega C_\alpha(\omega),
\end{equation}
It follows from (\ref{aratio}), (\ref{dratio}) and (\ref{lambda}) that
\begin{equation}
a(\omega)=\ln\frac{C_3^4(\omega)}{C_1(\omega)C_2^{\frac{1}{2}}(\omega)}
=4\ln C_3(\omega)-\ln C_1(\omega)-\frac{1}{2}\ln C_2(\omega)
\label{a0}
\end{equation}
The asymptotic forms of $C_1(\omega)$, $C_2(\omega)$ and $C_3(\omega)$ for large $\omega$ and small $\omega$ can be determined analytically. We find
\begin{equation}
\ln C_1(\omega)\simeq {\frac{\Gamma^2(\frac{1}{4})}{12\sqrt{\pi}\omega}}\simeq \frac{0.61802}{\omega}\qquad
\ln C_2(\omega)\simeq {\frac{5\Gamma^2(\frac{1}{4})}{6\sqrt{\pi}\omega}}\simeq\frac{6.1802}{\omega}\qquad
\ln C_3(\omega)\simeq {\frac{\Gamma^2(\frac{1}{4})}{8\sqrt{\pi}\omega}}\simeq \frac{0.92704}{\omega},
\label{clarge}
\end{equation}
as $\omega\to\infty$ and
\begin{equation}
C_1(\omega)\simeq\frac{\Gamma^2\left(\frac{1}{4}\right)}{8\sqrt{\pi}}\simeq\frac{0.927}{\omega}\qquad
C_2(\omega)\simeq\frac{225\Gamma^2\left(\frac{1}{4}\right)}{64\sqrt{\pi}}\simeq\frac{26.0736}{\omega^5}\qquad
C_3(\omega)\simeq\frac{1}{2\omega^2}
\label{csmall}
\end{equation}
as $\omega\to 0$.
The details of the derivation are shown in the appendices B and C. It follows from (\ref{a0}) that the integration over $\omega$ is convergent.

\subsection{The formulae of $b(\omega)$}

With a nonzero $\epsilon$, the Gelfand-Yaglom method becomes complicated as the differential equations of $\xi$ and $\eta$ are coupled, but we could employ the perturbation theory to calculate $b(\omega)$.

From (\ref{z}), we have
\begin{eqnarray}
\ln Z(\omega)={2\ln\rm det {\cal D}_\pm(\omega)}+\frac{1}{2}\ln {\rm det} D_{\xi\eta}(\omega)-\frac{1}{2}\ln {\rm det} D_1(\omega)
-\frac{5}{2}\ln {\rm det} D_0(\omega)
\label{operater}
\end{eqnarray}
where ${\cal D}_\pm(\omega)$ is given by (\ref{det2}) and $D_{\xi\eta}(\omega), D_1(\omega)$, $D_0(\omega)$ are given by (\ref{det}). For convenience, we denote
\begin{eqnarray}
Z_{\xi\eta}(\omega)&=&-\frac{1}{2}\ln {\rm det}D_{\xi\eta}(\omega)=-\frac{1}{2}\ln {\rm det}[\tilde{D}_{\xi\eta}(\omega)+
\delta D_{\xi\eta}(\omega)]\nonumber\\
Z_\pm(\omega)&=&2\ln\rm det {\cal D}_\pm(\omega)=2\ln {\rm det}[\tilde{{\cal D}}_\pm(\omega)+\delta{\cal D}_\pm(\omega) ]
\end{eqnarray}
where
\begin{eqnarray}
\tilde{D}_{\xi\eta}(\omega)=\left(\begin{array}{cc} -D_2(\omega) & 0 \\ 0 & D_1(\omega) \end{array}\right)\qquad
\delta D_{\xi\eta}(\omega)=\left(\begin{array}{cc} 0 & 4i\epsilon\omega e^\phi \\ -4i\epsilon\omega e^\phi & 0 \end{array}\right)\nonumber\\
\tilde{{\cal D}}_\pm(\omega)=\left(\begin{array}{cc} D_3(\omega) & 0 \\ 0 & D_3'(\omega)\qquad \end{array}\right)\qquad
\delta{\cal D}_\pm(\omega)=\left(\begin{array}{cc} \mp2i\epsilon\omega e^\phi & 0 \\ 0 & \mp2i\epsilon\omega e^\phi \end{array}\right)
\end{eqnarray}
with $D_3(\omega)$ and $D_3'(\omega)$ given by (\ref{det1}) and $D_2(\omega)$ given by (\ref{det3}). The next step is to expand $Z_{\xi\eta}(\omega)$ and $Z_\pm(\omega)$ to
the 2nd order in $\delta D_{\xi\eta}(\omega)$ and $\delta{\cal D}_\pm(\omega)$. The 1st order terms do not contribute because
${\rm tr}\delta D_{\xi\eta}(\omega)={\rm tr}\delta{\cal D}_\pm(\omega)=0$.
Consider first the $Z_{\xi\eta}(\omega)$
\begin{eqnarray}
Z_{\xi\eta}(\omega)&=&-\frac{1}{2}\ln {\rm det}D_{\xi\eta}(\omega)\nonumber\\
&=&-\frac{1}{2}\ln {\rm det}[\tilde{D}_{\xi\eta}(\omega)+\delta D_{\xi\eta}(\omega)]\nonumber\\
&=&-\frac{1}{2}\ln {\rm det}[\tilde{D}_{\xi\eta}(\omega)+\ln (1+\delta \tilde{D}^{-1}_{\xi\eta}(\omega)\delta D_{\xi\eta}(\omega))]\nonumber\\
&=&-\frac{1}{2}\ln {\rm det}\tilde{D}_{\xi\eta}(\omega)-\frac{1}{2}\textrm{Tr}\tilde{D}^{-1}_{\xi\eta}(\omega)\delta D_{\xi\eta}(\omega)+\frac{1}{4}\textrm{Tr}\tilde{D}^{-1}_{\xi\eta}(\omega)\delta D_{\xi\eta}(\omega)\tilde{D}^{-1}_{\xi\eta}(\omega)\delta D_{\xi\eta}(\omega)\nonumber\\
&=&-\frac{1}{2}\ln {\rm det}\tilde{D}_{\xi\eta}(\omega)+\frac{1}{4}\textrm{Tr}\tilde{D}^{-1}_{\xi\eta}(\omega)\delta D_{\xi\eta}(\omega)\tilde{D}^{-1}_{\xi\eta}(\omega)\delta D_{\xi\eta}(\omega)\nonumber\\
&=&\tilde{Z}_{\xi\eta}(\omega)+\delta\tilde{Z}_{\xi\eta}(\omega).
\end{eqnarray}
Likewise,
\begin{eqnarray}
Z_\pm(\omega)=\tilde{Z}_\pm(\omega)+\delta\tilde{Z}_\pm(\omega).
\end{eqnarray}
One can readily verify that at $\epsilon\neq0$
\begin{eqnarray}
\ln Z(\omega)=a(\omega)+\delta\tilde{Z}_{\xi\eta}(\omega)+\delta\tilde{Z}_\pm(\omega).
\end{eqnarray}
where $a(\omega)$ has been discussed in the previous section.
It then follows from (\ref{exp}) that
\begin{equation}
\epsilon^2b(\omega)=\delta\tilde{Z}_{\xi\eta}(\omega)+\delta\tilde{Z}_\pm(\omega)
\label{bb}
\end{equation}
where
\begin{eqnarray}
\delta\tilde{Z}_{\xi\eta}(\omega)&=&\frac{1}{4}\textrm{Tr}\tilde{D}^{-1}_{\xi\eta}(\omega)\delta D_{\xi\eta}(\omega)
\tilde{D}^{-1}_{\xi\eta}(\omega)\delta D_{\xi\eta}(\omega)\nonumber\\
&=&\frac{1}{4}\textrm{Tr}\left(\begin{array}{cc} -D_2^{-1}(\omega) & 0 \\ 0 & D_1^{-1}(\omega) \end{array}\right)
\left(\begin{array}{cc} 0 & 4i\epsilon\omega e^\phi \\ -4i\epsilon\omega e^\phi & 0 \end{array}\right)\left(\begin{array}{cc}
-D_2^{-1}(\omega) & 0 \\ 0 & D_1^{-1}(\omega) \end{array}\right)
\left(\begin{array}{cc} 0 & 4i\epsilon\omega e^\phi \\ -4i\epsilon\omega e^\phi & 0 \end{array}\right)\nonumber\\
&=&-8\epsilon^2\omega^2\textrm{Tr}D_1(\omega)^{-1}e^\phi D_2(\omega)^{-1}e^\phi\nonumber\\
&=&-16\epsilon^2\omega^2\int_{-\infty}^\infty d\rho\int_{-\infty}^\rho d\rho'g_{1\omega}(\rho-\rho')e^{\phi(\rho')}
g_{2\omega}(\rho-\rho')e^{\phi(\rho)}
\label{duv}
\end{eqnarray}
where the Green functions $g_{1\omega}(\rho-\rho')$, $g_{2\omega}(\rho-\rho')$ satisfy
\begin{eqnarray}
(-\frac{d^2}{d\rho^2}+\omega^2+e^{2\phi} M_1^2)g_{1\omega}(\rho-\rho')=\delta(\rho-\rho')\nonumber\\
(-\frac{d^2}{d\rho^2}+\omega^2+e^{2\phi} M_2^2)g_{2\omega}(\rho-\rho')=\delta(\rho-\rho').
\end{eqnarray}
In terms of the solutions defined in (\ref{solution}), we find
\begin{eqnarray}
g_{1\omega}(\rho-\rho')=\frac{1}{2\omega C_1(\omega)}u_{1\omega}^{(1)}(\rho')u_{1\omega}^{(2)}(\rho)
\label{g1}
\end{eqnarray}
\begin{eqnarray}
g_{2\omega}(\rho-\rho')=\frac{1}{2\omega C_2(\omega)}u_{2\omega}^{(1)}(\rho')u_{2\omega}^{(2)}(\rho)
\label{g2}
\end{eqnarray}
for $\rho'<\rho$, where the dependence on $\omega$ is indicated explicitly. In what follows, we shall suppress the
superscripts and denote $u_{\alpha\omega}^{(1)}(\rho)$ by $u_{\alpha\omega}(\rho)$ for $\alpha=1,2$.
It follows from the symmetry of their differential equations that $u_{\alpha\omega}^{(2)}(\rho)=u_{\alpha\omega}(-\rho)$.

Substituting (\ref{g1}), (\ref{g2}) into (\ref{duv}), we have
\begin{eqnarray}
\delta\tilde{Z}_{\xi\eta}(\omega)=-\frac{4\epsilon^2}{C_1(\omega)C_2(\omega)}\int_{-\infty}^\infty d\rho e^{\phi(\rho)} u_{1\omega}
(-\rho)u_{2\omega}(-\rho)\int_{-\infty}^\rho d\rho' e^{\phi(\rho')} u_{1\omega}
(\rho')u_{2\omega}(\rho').
\label{duv1}
\end{eqnarray}
The process to calculate $\delta\tilde{Z}_\pm(\omega)$ is similar to the $\delta\tilde{Z}_{\xi\eta}(\omega)$ part, here we just show the result
\begin{eqnarray}
\delta\tilde{Z}_\pm(\omega)=\frac{4\epsilon^2}{C_3^2(\omega)}\int_{-\infty}^\infty d\rho e^{\phi(\rho)} v_{3\omega}^2
(\rho)\int_{-\infty}^\rho d\rho' e^{\phi(\rho')} u_{3\omega}^2
(\rho').
\label{df1}
\end{eqnarray}
where $u_{3\omega}(\rho)$ and $v_{3\omega}(\rho)$ stand for $u_3^{(1)}(\rho)$ and $u_3^{(2)}(\rho)$ defined in (\ref{solution}).
Applying (\ref{duv1}),(\ref{df1}) into (\ref{bb}), we find
\begin{eqnarray}
b(\omega)&=&-\frac{4}{C_1(\omega)C_2(\omega)}\int_{-\infty}^\infty d\rho e^{\phi(\rho)} u_{1\omega}
(-\rho)u_{2\omega}(-\rho)\int_{-\infty}^\rho d\rho' e^{\phi(\rho')} u_{1\omega}
(\rho')u_{2\omega}(\rho')\nonumber\\
&+&\frac{4}{C_3^2(\omega)}\int_{-\infty}^\infty d\rho e^{\phi(\rho)} v_{3\omega}^2
(\rho)\int_{-\infty}^\rho d\rho' e^{\phi(\rho')} u_{3\omega}^2
(\rho').
\label{b0}
\end{eqnarray}
To prove the convergence of the integral of $b(\omega)$ over $\omega$, let us examine the large $\omega$ behavior
first. The solutions of $D_{1,2,3}(\omega)u_{1,2,3}=0$ may be approximated by $u_{1,2,3}(\rho)\simeq e^{\omega \rho}$ and
$C_{1,2,3}=1+O\left(\frac{1}{\omega}\right)$. Introducing the new variables $X=\frac{1}{2}(\rho+\rho')$ and
$x=\rho-\rho'$, We have then
\begin{eqnarray}
&&-\frac{4}{C_1(\omega)C_2(\omega)}\int_{-\infty}^\infty d\rho e^{\phi(\rho)} u_{1\omega}
(-\rho)u_{2\omega}(-\rho)\int_{-\infty}^\rho d\rho' e^{\phi(\rho')} u_{1\omega}
(\rho')u_{2\omega}(\rho')\nonumber\\
&=&\int_{-\infty}^\infty dX\int_0^\infty dx
e^{\phi\left(X+\frac{x}{2}\right)+\phi\left(X-\frac{x}{2}\right)-2\omega x}
=-\frac{2}{\omega}\int_{-\infty}^\infty dXe^{2\phi(X)}+o\left(\frac{1}{\omega}\right)
\end{eqnarray}
and similarly
\begin{eqnarray}
&&\frac{4}{C_3^2(\omega)}\int_{-\infty}^\infty d\rho e^{\phi(\rho)} v_{3\omega}^2
(\rho)\int_{-\infty}^\rho d\rho' e^{\phi(\rho')} u_{3\omega}^2(\rho')\nonumber\\
&=&\frac{2}{\omega}\int_{-\infty}^\infty dXe^{2\phi(X)}+o\left(\frac{1}{\omega}\right).
\end{eqnarray}
Consequently, the leading terms of the two integral in (\ref{b0}) cancel and $b(\omega)$ vanishes faster than
$O\left(\frac{1}{\omega}\right)$ as $\omega\to\infty$.

The integral of $b(\omega)$ is also convergent at the lower limit because
\begin{eqnarray}
\lim_{\omega\to 0}b(\omega)\simeq 2.
\end{eqnarray}
the details are shown in appendix D.

\section{The numerical calculations}
In this section, we present the numerical calculations of $a(\omega)$ and $b(\omega)$.

Consider the equation
\begin{equation}
\frac{d^2u}{d\rho^2}-(\omega^2+V)u=0.
\label{equation1}
\end{equation}
where V refers to the potential of the Schrodinger like equation $D_iu=0,(i=1,2,3)$.
To avoid the exponentially growing behavior, we factor out $e^{\omega\rho}$ and set $u(\rho)=\eta(\rho)e^{\omega\rho}$, then the equation (\ref{equation1}) becomes
\begin{equation}
\frac{d^2\eta}{d\rho^2}+2\omega\frac{d\eta}{d\rho}-V\eta=0.
\label{equation2}
\end{equation}
which is suitable for the numerical calculations, then coefficients, $C_1(\omega)$, $C_2(\omega)$, $C_3(\omega)$
together with the solutions $\eta_i(\rho)$ can be obtained from the equation (\ref{equation2}). This method has been employed recently in Ref \cite{chu:2009} and \cite{zhang2011}.

We start from the negative value of $\rho$ with $|\rho|>>1$, where these Schrodinger like equations take the asymptotic forms
\begin{eqnarray}
D_1u_1&\simeq&(-\frac{d^2}{d\rho^2}+\omega^2)u_1=0\nonumber\\
D_2u_2&\simeq&(-\frac{d^2}{d\rho^2}+\omega^2+\frac{6}{\rho^2})u_2=0\nonumber\\
D_3u_3&\simeq&(-\frac{d^2}{d\rho^2}+\omega^2+\frac{2}{\rho^2})u_3=0
\label{D122}
\end{eqnarray}
At $\rho=-K$ ($K>>1$ is a large cutoff, and $K\rightarrow\infty$).
We can find the solutions of (\ref{D122})
\begin{eqnarray}
u_1(-K)&=&e^{-\omega K}\nonumber\\
u_2(-K)&=&(1+\frac{3}{\omega K}+\frac{3}{\omega^2K^2})e^{-\omega K}\nonumber\\
u_3(-K)&=&(1+\frac{1}{\omega K})e^{-\omega K},
\end{eqnarray}
It follows from $\eta(\rho)=e^{-\omega\rho}u(\rho)$ that
\begin{eqnarray}
\eta_1(-K)&=&1,\qquad \eta'_1(-K)=0\nonumber\\
\eta_2(-K)&=&1+\frac{3}{\omega K}+\frac{3}{\omega^2K^2}, \qquad\eta'_2(-K)=\frac{3}{\omega K^2}+\frac{6}{\omega^2K^3}\nonumber\\
\eta_3(-K)&=&1+\frac{1}{\omega K},\qquad \eta'_3(-K)=\frac{1}{\omega K^2}
\end{eqnarray}
which serve as the initial conditions for the numerical solutions. Then we run the forth order Runge-Kutta algorithm
all the way to $\rho=K$. There, we can find the approximate solutions as well (See appendix C for details), from which
$C_1(\omega)$, $C_2(\omega)$ and $C_3(\omega)$ are extracted with the formula
\begin{eqnarray}
C_1(\omega)&=&\eta_1(K)+\frac{\eta'_1(K)}{2\omega}\nonumber\\
C_2(\omega)&=&\frac{\eta_2(K)N+[\eta'_2(K)+\omega\eta_2(K)]Q}{2\omega}\nonumber\\
C_3(\omega)&=&\eta_3(K)+\frac{\eta'_3(K)}{2\omega}
\end{eqnarray}
where
\begin{eqnarray}
N&=&\omega+\frac{3}{K}+\frac{6}{\omega K^2}+\frac{6}{\omega^2 K^3}\nonumber\\
Q&=&1+\frac{3}{\omega K}+\frac{3}{\omega^2 K^2}.
\end{eqnarray}
The numerically generated $C_1(\omega)$, $C_2(\omega)$ and $C_3(\omega)$ for large $\omega$ and small $\omega$ read
\begin{equation}
\ln C_1(\omega)=\frac{0.61799}{\omega}\qquad
\ln C_2(\omega)=\frac{6.1799}{\omega}\qquad
\ln C_3(\omega)=\frac{0.92703}{\omega}
\end{equation}
\begin{equation}
C_1(\omega)=\frac{0.927}{\omega}\qquad
C_2(\omega)=\frac{26.0971}{\omega^5}\qquad
C_3(\omega)=\frac{0.5001}{\omega^2}.
\end{equation}
Comparing the above behaviors with (\ref{clarge}) and (\ref{csmall}), we find that the numerical results are in agreement
with the analytical ones. Coming back to (\ref{k}), owning to the evenness in $\omega$, we have
\begin{eqnarray}
\kappa=\frac{8\sqrt{\pi}}{\Gamma^2{(\frac{1}{4})}}\int_{-\infty}^\infty d\omega
\Big[\frac{1}{2}a(\omega)-b(\omega)\Big]=\frac{16\sqrt{\pi}}{\Gamma^2{(\frac{1}{4})}}\int_0^\infty
d\omega\Big[\frac{1}{2}a(\omega)-b(\omega)\Big].
\label{k1}
\end{eqnarray}
Having obtained the numerical values of $C_1(\omega),C_2(\omega),C_3(\omega)$, the integration of $\frac{1}{2}a(\omega)$
is carried out in the following steps. The Simpson rule yields
\begin{equation}
\int_{0.001}^{14}d\omega
\frac{1}{2}a(\omega)=1.548.
\label{a1}
\end{equation}
Considering the small $\omega$ behavior of $C_\alpha$ from (\ref{csmall}), we can fit the expression of $\frac{1}{2}a(\omega)$
\begin{eqnarray}
\frac{1}{2}a(\omega)\simeq-2.155-2.249\ln(\omega)
\end{eqnarray}
as $\omega\rightarrow0$. Thus
\begin{eqnarray}
\int_{0}^{0.001}d\omega
\frac{1}{2}a(\omega)\simeq\int_{0}^{0.001}d\omega[-2.155-2.249\ln(\omega)]=0.016.
\label{a2}
\end{eqnarray}
Likewise, the expression of $\frac{1}{2}a(\omega)$ when $\omega$ is large can be fitted to
\begin{eqnarray}
\frac{1}{2}a(\omega)\simeq\frac{0.002}{\omega^2}+\frac{0.646}{\omega^3}
\end{eqnarray}
then
\begin{eqnarray}
\int_{14}^{\infty}d\omega
\frac{1}{2}a(\omega)\simeq\int_{14}^{\infty}d\omega[\frac{0.002}{\omega^2}+\frac{0.646}{\omega^3}]
=0.002
\label{a3}
\end{eqnarray}
Combining (\ref{a1}), (\ref{a2}) and (\ref{a3}), we have
\begin{equation}
\int_{0}^{\infty}d\omega
\frac{1}{2}a(\omega)=1.566.
\label{a}
\end{equation}

The function $b(\omega)$ can be divided into two parts
\begin{eqnarray}
b(\omega)&=&-\frac{4}{C_1(\omega)C_2(\omega)}\int_{-\infty}^\infty d\rho e^{\phi(\rho)} u_{1\omega}
(-\rho)u_{2\omega}(-\rho)\int_{-\infty}^\rho d\rho' e^{\phi(\rho')} u_{1\omega}
(\rho')u_{2\omega}(\rho')\nonumber\\
&+&\frac{4}{C_3^2(\omega)}\int_{-\infty}^\infty d\rho e^{\phi(\rho)} v_{3\omega}^2
(\rho)\int_{-\infty}^\rho d\rho' e^{\phi(\rho')} u_{3\omega}^2
(\rho')\nonumber\\
&=& b_1(\omega)+b_2(\omega).
\label{b1}
\end{eqnarray}
where $b_1(\omega)$ and $b_2(\omega)$ are the 1st and the 2nd double integrals on RHS.
In terms of the numerical values of $C_1(\omega), C_2(\omega)$ and $C_3(\omega)$ and the corresponding solutions $\eta_{i\omega}(\rho)$ at each $\omega$,
the numerical values of $b_1(\omega)$ and $b_2(\omega)$ can be obtained.
Combining $b_1(\omega)$ and $b_2(\omega)$, the curve of $b(\omega)$ is shown in FIG.2.
\begin{figure}
\centering
\includegraphics[height=2.0in]{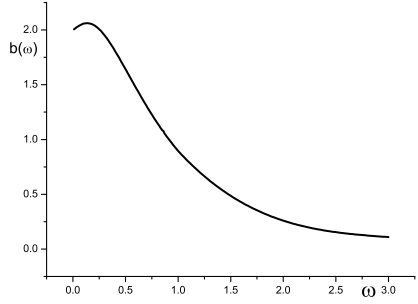}
\caption{The curve of $b(\omega)$ vs $\omega$.}
\end{figure}
Notice that the integral of $b(\omega)$ over $\omega$ is convergent since $b(\omega)$ falls off faster than $\omega^{-1}$
towards the upper limit and approaches a finite limit at lower limit, consistent with the analytical results of last section.
When $\omega$ is not large, we compute the integral by means of the trapezoid method and obtain
\begin{eqnarray}
\int_{0}^{3} d\omega b(\omega)\simeq2.283.
\label{b1}
\end{eqnarray}
For large $\omega$, $b(\omega)$ can be fitted to
\begin{eqnarray}
b(\omega)\simeq\frac{0.21}{\omega^2}+\frac{2.329}{\omega^3}
\end{eqnarray}
and the integral is carried out analytically. We find
\begin{eqnarray}
\int_3^\infty d\omega b(\omega)\simeq\int_3^\infty d\omega[\frac{0.21}{\omega^2}+\frac{2.329}{\omega^3}]=0.197
\label{b2}
\end{eqnarray}

Combining (\ref{b1}) and (\ref{b2}), we have
\begin{equation}
\int_{0}^\infty d\omega
b(\omega)=2.283+0.197=2.48.
\label{b22}
\end{equation}
From (\ref{a}) and (\ref{b22}), we find
\begin{eqnarray}
 \int_{0}^\infty d\omega
[\frac{1}{2}a(\omega)-b(\omega)]=1.566-2.48=-0.914.
\label{inter1}
\end{eqnarray}
Substituting (\ref{inter1}) into (\ref{k1}), we end up with the numerical value of $\kappa$
\begin{eqnarray}
 \kappa=\frac{16\sqrt{\pi}}{\Gamma^2{(\frac{1}{4})}}\int_0^\infty d\omega
[\frac{1}{2}a(\omega)-b(\omega)] \simeq -1.97.
\end{eqnarray}
The physical significance of $\kappa$ will be discussed in the next section.

\section{Concluding remarks}

The jet quenching is an important signal of the quark-gluon plasma created via heavy ion collisions.
At the energy scale of RHIC, the QCD running coupling constant is not weak, the perturbative
analysis may not be appropriate, nor the lattice is available. In terms
of the QCD running coupling constant $g_{\rm QCD}$ and $\alpha_{\rm QCD}=\frac{g_{\rm QCD}^2}{4\pi}$,
the 't Hooft coupling $\lambda_{\rm QCD}=4\pi N_c\alpha_{\rm QCD}=12\pi\alpha_{\rm QCD}$ and is strong
when $\alpha_{\rm QCD}=O(1)$. This motivates the application of AdS/CFT to the $\mathcal N=4$ super Yang-Mills
plasma as an important reference. To the leading order of large
$N_c$ and strong 't Hooft coupling $\lambda$, the jet quenching parameter in the super Yang-Mills plasma has been calculated. The sub-leading term in the strong coupling expansion, which dominates the $O(\lambda^{-\frac{3}{2}})$
term in the literature, is provided in this paper.

To figure out the leading order correction, we have expanded the Nambu-Goto action around the word sheet
underlying the order jet quenching in the quadratic order in the fluctuation coordinates and carried out.
As the leading order world sheet is a saddle point of the Nambu-Goto action,
we defined the one loop effective action as an analytical continuation from a path integral with
an imaginary string tension, from which the correction is extracted. The resultant effective action is expressed
in terms of ratios of determinants and is shown explicitly to be divergence free. The reason for the absence of UV
divergence can be attributed to the the metric (\ref{target}) solves the 10d supergravity equation of motion \cite{Townsend}.
The numerical computation of the determinant ratios yields the result (\ref{result}). The negative sign of $\kappa$ is consistent a
monotonic behavior from strong coupling to weak coupling. For $\alpha_{SYM}=1/2$ and $N_c=3$, we have
$\lambda=18.8$, which gives rise to 32\% reduction of $\hat q_{SYM}$ from the leading order amount.

The jet quenching parameter has also been calculated within the AdS/CFT correspondence via the dragging force
with a slightly different outcome to the leading order. As this calculation also involves the minimum area
under the Nambu-Goto action, it remains to examine the order $\lambda^{-\frac{1}{2}}$ along this line. This
discrepancy from the experimental result may also stem from other difference between QCD and the super Yang-Mills
at large $N_c$ and large 't Hooft coupling, such as finite $N_c$ and infrared cutoff of the former. We hope
to report our progress in this regard.

\acknowledgments
  We thank J. X.  Lu for valuable discussions. The research of Defu Hou and Hai-cang Ren is  is  supported partly by NSFC under grant Nos. 10975060, 11135011, 11221504. The work  of  Ziqiang Zhang is supported in part by the QLPL under grant Nos. QLPL201113.

\appendix
\section {The UV divergence of the bosonic and fermionic fluctuations}

In this appendix, we shall prove the cancelation of the UV divergence caused by the mass terms of the
fluctuation action (including the cross term of (\ref{xieta})), which contributes to $c_1$ of (\ref{uvmass}).
The world sheet curvature does not contribute in this case and we may work in an cartesian frame in the
neighborhood of each point, $\sigma_0\equiv(r_0,\tau_0)$. To make all path integrals meaningful, we shall
start with an imaginary $\alpha'$ in accordance with the discussion following (\ref{partition}).

\noindent
1. {\it The $\zeta$ mode}

\begin{equation}
\hat S_{\zeta}=\frac{i}{2}\int d\tau dr\sqrt{\bar{g}}{(\bar{g} ^{\tau\tau}\dot{\zeta} ^2+
\bar{g} ^{rr}{\zeta^{'}}^2+M_\zeta^2\zeta^2})
\end{equation}
Define
\begin{eqnarray}
r-r_0=\frac{1}{\sqrt{\bar{g}_{rr}}}x_1\nonumber\\
\tau-\tau_0=\frac{1}{\sqrt{\bar{g}_{\tau\tau}}}x_2
\end{eqnarray}
then
\begin{eqnarray}
ds^2=dx_1^2+dx_2^2\nonumber\\
\int d\tau\int dr\sqrt{\bar{g}}=\int d^2x
\end{eqnarray}
and
\begin{equation}
\hat S_\zeta=\sum_{\sigma_0}\Delta \hat S_\zeta
\end{equation}
where
\begin{equation}
\Delta \hat S_\zeta=\frac{i}{2}\int d^2x(\partial_\alpha\zeta\partial_\alpha\zeta+M_\zeta^2\zeta^2)
\end{equation}
Making a Fourier expansion
\begin{eqnarray}
\zeta&=&\frac{1}{\sqrt{\Omega}}\sum_{\vec{p}}\zeta_{\vec{p}}e^{i\vec{p}\vec{r}}\nonumber\\
\zeta_{\vec{p}}^\ast&=&\zeta_{-{\vec{p}}}
\end{eqnarray}
where $\Omega$ is the area covered by the cartesian frame at $\sigma_0$, then
\begin{equation}
\Delta \hat S_\zeta=\frac{i}{2}\sum_{\vec{p}}(\vec{p}^2+M_\zeta^2)\zeta_{\vec{p}}^\ast\zeta_{\vec{p}}=
{\sum_{\vec{p}}}'(\vec{p}^2+M_\zeta^2)\zeta_{\vec{p}}^\ast\zeta_{\vec{p}}
\end{equation}
with ${\sum_{\vec{p}}}'$ extending to half momentum space.

On writing
\begin{equation}
\int{\prod_{\vec{p}}}'d\zeta_{\vec{p}}^*d\zeta_{\vec{p}}e^{\frac{i}{2\pi\alpha'}\Delta \hat S_\zeta}
\equiv {\rm const.}e^{-\Delta\Gamma_\zeta},
\end{equation}
where
\begin{equation}
\Delta\Gamma_\zeta=\frac{1}{2}\sum_{\vec{p}}\ln(\vec{p}^2+M_\zeta^2)=\frac{1}{2}\sum_{\vec{p}}\ln\vec{p}^2+
\frac{1}{2}M_\zeta^2\sum_{\vec{p}}\frac{1}{p^2}
\end{equation}
and the constant does not contribute to $\hat q$. So the mass induced logarithmic divergence
\begin{eqnarray}
({\rm UV})_\zeta=\frac{1}{2}M_\zeta^2\sum_{p<\Lambda}\frac{1}{p^2}\simeq\frac{\Omega}{4\pi}M_\zeta^2\int^{\Lambda}\frac{dp}{p}\simeq
\frac{\Omega}{4\pi}M_\zeta^2\ln\Lambda
\end{eqnarray}

\noindent
2. {\it The $\xi$ and $\eta$ modes}

\begin{eqnarray}
S_{\xi\eta}&=&\frac{i}{2}\int d\tau dr\sqrt{\bar{g}}(-\bar{g} ^{\tau\tau}\dot{\xi} ^2-
\bar{g} ^{rr}{\xi'}^2+\bar{g} ^{\tau\tau}\dot{\eta} ^2+
\bar{g} ^{rr}{\eta'}^2-M_\xi^2\xi^2+M_\eta^2\eta^2-\frac{8\sqrt{2}\epsilon}{\sqrt{1+\epsilon^2}}\frac{r}{r_h^2}\xi\dot{\eta})
\nonumber\\
&=&\sum_{\sigma_0}\Delta \hat S_{\xi\eta}
\end{eqnarray}
where
\begin{equation}
\Delta \hat S_{\xi\eta}=\frac{i}{2}\int d^2x(-\partial_\alpha\xi\partial_\alpha\xi+\partial_\alpha\eta\partial_\alpha\eta+
-M_\xi^2\xi^2+M_\eta^2\eta^2-\frac{8\sqrt{2}\epsilon}{\sqrt{1+\epsilon^2}}\xi\partial_2\eta)
\end{equation}
Making a Fourier transformation
\begin{eqnarray}
\xi=\frac{1}{\sqrt{\Omega}}\sum_{\vec{p}}\xi_{\vec{p}}e^{i\vec{p}\vec{r}}\nonumber\\
\eta=\frac{1}{\sqrt{\Omega}}\sum_{\vec{p}}\eta_{\vec{p}}e^{i\vec{p}\vec{r}}
\end{eqnarray}
then
\begin{eqnarray}
\hat S_{\xi\eta}=\frac{i}{2}\sum_{\vec{p}}[-(\vec{p}^2+M_\xi^2)\xi_{\vec{p}}^\ast\xi_{\vec{p}}+(\vec{p}^2+
M_\eta^2)\eta_{\vec{p}}^\ast\eta_{\vec{p}}-\frac{8i\epsilon}{\sqrt{1+\epsilon^2}}\xi_{\vec{p}}^\ast p_2\eta_{\vec{p}}]\nonumber\\
{\sum_{\vec{p}}}'[(\vec{p}^2+M_\xi^2)\xi_{\vec{p}}^\ast\xi_{\vec{p}}+(\vec{p}^2+
M_\eta^2)\eta_{\vec{p}}^\ast\eta_{\vec{p}}+\frac{4\epsilon}{\sqrt{1+\epsilon^2}}p_2(\xi_{\vec{p}}^\ast\eta_p-
\eta_{\vec{p}}^\ast\xi_p)]
\end{eqnarray}
On writing
\begin{equation}
\int{\prod_{\vec{p}}}'d\xi_{\vec{p}}^*d\xi_{\vec{p}}d\eta_{\vec{p}}^*d\eta_{\vec{p}}
e^{\frac{i}{2\pi\alpha'}\Delta \hat S'_{\xi\eta}} \equiv {\rm const.}e^{-\Delta\Gamma_{\xi\eta}}
\end{equation}
we find
\begin{eqnarray}
\Delta\Gamma_{\xi\eta} &=& \ln{\prod_{\vec{p}}}'\left|\begin{array}{cc} p^2+M_\xi^2 & -\frac{4i\epsilon p_2}{\sqrt{1+\epsilon^2}}
\\ -\frac{4i\epsilon p_2}{\sqrt{1+\epsilon^2}} & p^2+M_\eta^2 \end{array}\right|\nonumber\\
&=& \frac{1}{2}\sum_{\vec{p}}[2\ln p^2+
\frac{1}{p^2}(M_\varepsilon^2+M_\eta^2+\frac{8\sqrt{2}\epsilon}{\sqrt{1+\epsilon^2}})+...].
\end{eqnarray}
Then the mass induced logarithmic divergence
\begin{eqnarray}
({\rm UV})_{\xi\eta}=\frac{1}{2}(M_\xi^2+M_\eta^2+\frac{8\sqrt{2}\epsilon}{\sqrt{1+\epsilon^2}})
\sum_{p<\Lambda}\frac{1}{p^2}\simeq\frac{\Omega}{4\pi}(M_\xi^2+M_\eta^2+\frac{8\sqrt{2}\epsilon}{\sqrt{1+\epsilon^2}})
\ln\Lambda
\end{eqnarray}
It follows that the coefficient of the total logarithmic divergence due to the mass terms of the bosonic fluctuations
\begin{eqnarray}
\frac{1}{2}\int d^2\sigma\sqrt{\bar g}
\left(M_\zeta^2+M_\xi^2+M_\eta^2+\frac{8\sqrt{2}\epsilon}{\sqrt{1+\epsilon^2}}\right)\nonumber\\
=\frac{1}{2}\int d^2\sigma\sqrt{\bar g}(8+R)=4\int d^2\sigma\sqrt{\bar g}
\end{eqnarray}

\noindent
3. {\it Fermionic modes}

The operator (\ref{wfermion})
can be related to a massive Dirac operator of in a 2D curved space and a $U(1)$ gauge potential, i.e.
\begin{eqnarray}
{\cal A}_\pm = i\sigma_2\left(\gamma^\alpha\nabla_\alpha+m\right),
\end{eqnarray}
where the mass $m=1$ and the covariant derivative
\begin{equation}
\nabla_\alpha=\frac{\partial}{\partial\sigma^\alpha}-iA_\alpha+\frac{1}{4}[\gamma,\gamma_b]\omega_\alpha^{ab}
\end{equation}
with the 2D gamma matrices $\gamma^0=\sigma_1$, $\gamma^1=-\sigma_3$ and the gauge potential
$A_\tau=\pm i\frac{\epsilon}{\sqrt{1+\epsilon^2}}e^0_\tau$, $A_r=0$. As is in a flat 2D space,
the gauge potential does not contribute to the UV divergence. Therefore, the UV divergence of the
fermionic effective action in (\ref{partition})
\begin{eqnarray}
\Gamma_F=2(\ln{\rm det}{\cal A}_++\ln{\rm det}{\cal A}_-)
\end{eqnarray}
is the same as
\begin{eqnarray}
\Gamma_F^\prime\equiv 4\ln{\rm det}{\cal A}=2\ln{\rm det}{\cal A}^2,
\end{eqnarray}
where ${\cal A}$ is obtained from ${\cal A}_\pm$ with the covariant derivative replaced by
\begin{equation}
\nabla_\alpha=\frac{\partial}{\partial\sigma^\alpha}+\frac{1}{4}[\gamma,\gamma_b]\omega_\alpha^{ab}.
\end{equation}
We have
\begin{eqnarray}
{\cal A}^2=\nabla^\alpha\nabla_\alpha-1-\frac{1}{4}R.
\end{eqnarray}
It follows that the coefficient of the logarithmic divergence is given by
\begin{equation}
\int d^2\sigma\sqrt{\bar g}(4+{\rm const}\times R)=4\int d^2\sigma\sqrt{\bar g},
\end{equation}
which cancel the bosonic contribution.

\section {The large $\omega$ behavior}
In this appendix, we present the details of the derivation of
$C_1(\omega), C_2(\omega), C_3(\omega)$ at a large and positive $\omega$.\\

We start with the 2nd order ordinary differential equations
\begin{equation}
(-\frac{d^2}{d\rho^2}+\omega^2+e^{2\phi}M_1^2)f_1=0,
\end{equation}
\begin{equation}
(-\frac{d^2}{d\rho^2}+\omega^2+e^{2\phi}M_2^2)f_2=0.
\end{equation}
and
\begin{equation}
(-\frac{d^2}{d\rho^2}+\omega^2+e^{2\phi}-
e^\phi\frac{d\phi}{d\rho})f_3=0.
\end{equation}
Next, we find the WKB approximation large $\omega$
\begin{eqnarray}
f_1\cong e^{\pm\int^\rho
{d\rho^\prime\sqrt{\omega^2+e^{2\phi}M_1^2}}}\nonumber\\
f_2\cong e^{\pm\int^\rho
{d\rho^\prime\sqrt{\omega^2+e^{2\phi}M_2^2}}}
\end{eqnarray}
and
\begin{eqnarray}
f_3\cong e^{\pm\int^\rho{d\rho^\prime\sqrt{\omega^2+e^{2\phi}-
e^\phi\frac{d\phi}{d\rho}}}}
\end{eqnarray}. By choosing appropriate sign and the integration constant $\rho_0$ in each case, we may find the large $\omega$
approximations of the specific solutions $u_\alpha^{(1)}$ and $u_\alpha^{(2)}$ defined in (\ref{solution}).
In particular, we have
\begin{equation}
u_1^{(1)}(\rho)\cong e^{\omega\rho+\frac{1}{2\omega}\int^{\rho}_{-\infty}d\rho^\prime e^{2\phi}M_1^2},
\end{equation}
\begin{equation}
u_2^{(2)}(\rho)\cong e^{\omega\rho+\frac{1}{2\omega}\int^{\rho}_{-\infty}d\rho^\prime e^{2\phi}M_2^2}
\end{equation}
and
\begin{equation}
u_3^{(1)}(\rho)\cong e^{\omega\rho+\int^{\rho}_{-\infty}d\rho^\prime (e^{2\phi}- e^\phi\frac{d\phi}{d\rho^\prime})}
\end{equation}
Thus, we find that
\begin{eqnarray}
\ln C_1(\omega)&\cong& {\frac{1}{2\omega}\int^{\infty}_{-\infty}d\rho e^{2\phi}M_1^2}= {\frac{\Gamma^2(\frac{1}{4})}{12\sqrt{\pi}\omega}}\nonumber\\
\ln C_2(\omega)&\cong& {\frac{1}{2\omega}\int^{\infty}_{-\infty}d\rho e^{2\phi}M_2^2}= {\frac{5\Gamma^2(\frac{1}{4})}{6\sqrt{\pi}\omega}}\nonumber\\
\ln C_3(\omega)&\cong& {\frac{1}{2\omega}\int^{\infty}_{-\infty}d\rho (e^{2\phi}- e^\phi\frac{d\phi}{d\rho})}= {\frac{\Gamma^2(\frac{1}{4})}{8\sqrt{\pi}\omega}}.
\end{eqnarray}
as $\omega\to \infty$. The final results in terms of the Gamma functions are obtained by transforming the integral with respect to
$\rho$ back to that with respect to $r$ according to (\ref{rho}).

\section {The small $\omega$ behavior}
To estimate of the small $\omega$ behavior $C_1$, $C_2$ and $C_3$, we
introduce three domains of $\rho$, $I_{\rm R,L}:\hbox{}|\rho|<<1$ and $II:\hbox{}|\rho|<<1/\omega$
with the subscripts R and L referring to the right side $(\rho>0)$ and the left side $(\rho<0)$ of the
origin. For $\omega<<1$, the intersects $I_L\cap II$ and $II\cap I_R$ are nonempty and serves as
buffers where the analytically approximate solutions within different domains can be matched. In what follows, we
shall denote the set of functional operators involved by $D\equiv\lbrace D_1, D_2, D_3, D_3^\prime\rbrace$ and
their kernels by $u=\lbrace u_1, u_2, u_3, u_3^\prime\rbrace$. Though the
subset of $\lbrace D_1$, $D_2$, $D_3\rbrace$ is sufficient to determine $C_1(\omega)$, $C_2(\omega)$
and $C_3(\omega)$, we include $D_3^\prime$ in the early part of this appendix since $v_{3\omega}(\rho)$ of (\ref{solution})
can be obtained from the solution of $D_3^\prime u_3^\prime=0$ subject to the same boundary conditions of
$u_1$, $u_2$ and $u_3$ upon replacing $\rho$ by -$\rho$.

Within the domains $I_L$ and $I_R$, all powers $1/\rho$ in $D$ that fall off faster
than $1/\rho^2$ can be ignored and the equations $D\phi=0$ become
analytically soluble in terms of elementary functions. We have
\begin{eqnarray}
D_1&\simeq&-\frac{d^2}{d\rho^2}+\omega^2\nonumber\\
D_2&\simeq&-\frac{d^2}{d\rho^2}+\omega^2+\frac{6}{\rho^2}
\label{D12I}
\end{eqnarray}
for both $I_L$ and $I_R$, and
\begin{equation}
D_3\simeq-\frac{d^2}{d\rho^2}+\omega^2+\frac{2}{\rho^2}
\qquad D_3^\prime\simeq-\frac{d^2}{d\rho^2}+\omega^2
\label{D3IL}
\end{equation}
for $I_L$ and
\begin{eqnarray}
D_3\simeq-\frac{d^2}{d\rho^2}+\omega^2
\qquad D_3^\prime\simeq-\frac{d^2}{d\rho^2}+\omega^2+\frac{2}{\rho^2}
\label{D3IR}
\end{eqnarray}
for $I_R$. In the domain $II$, on the other hand, $\omega$ can be approximated by zero and $D$'s become
\begin{eqnarray}
D_1&\simeq&-\frac{d^2}{d\rho^2}+e^{2\phi}M_1^2\nonumber\\
D_2&\simeq&-\frac{d^2}{d\rho^2}+e^{2\phi}M_2^2\nonumber\\
D_3&\simeq&-\frac{d^2}{d\rho^2}+e^{2\phi}-e^\phi\frac{d\phi}{d\rho}\nonumber\\
D_3'&\simeq&-\frac{d^2}{d\rho^2}+e^{2\phi}+e^\phi\frac{d\phi}{d\rho}..
\label{DII}
\end{eqnarray}
In the overlapping regions $I_L\cap II$ and $II\cap I_R$
where both high powers of $\frac{1}{\rho}$ and $\omega$ can be dropped, we end up with
\begin{eqnarray}
D_1&\simeq&-\frac{d^2}{d\rho^2}\nonumber\\
D_2&\simeq&-\frac{d^2}{d\rho^2}+\frac{6}{\rho^2}
\label{DI_II}
\end{eqnarray}
for both $I_L\cap II$ and $II\cap I_R$, and
\begin{equation}
D_3\simeq-\frac{d^2}{d\rho^2}+\frac{2}{\rho^2} \qquad D_3^\prime\simeq-\frac{d^2}{d\rho^2}
\label{DI_II_L}
\end{equation}
within $I_L\cap II$ and
\begin{equation}
D_3\simeq-\frac{d^2}{d\rho^2} \qquad D_3^\prime\simeq-\frac{d^2}{d\rho^2}+\frac{2}{\rho^2}
\label{DI_II_R}
\end{equation}
within $II\cap I_R$.

In what follows, we shall
start with the approximate solutions in $I_L$ with the asymptotic form $e^{\omega\rho}$ as $\rho\to-\infty$,
matching different approximations through $II$ and $I_R$ and obtain the leading small $\omega$ behavior of
the coefficients of the asymptotic form $e^{\omega\rho}$ as $\rho\to\infty$.

The solutions of $Du=0$ in $I_L$ with $D$'s given by the approximation (\ref{D12I}) and (\ref{D3IL})
subject to the condition that
$u\to e^{\omega\rho}$ as $\rho\to -\infty$ read
\begin{eqnarray}
u_1&\simeq&e^{\omega\rho}\nonumber\\
u_2&\simeq&\sqrt{-\frac{2\omega\rho}{\pi}}K_{\frac{5}{2}}
(-\omega\rho)\nonumber\\
u_3&\simeq&\sqrt{-\frac{2\omega\rho}{\pi}}K_{\frac{3}{2}}(-\omega\rho)\nonumber\\
u_3'&\simeq&e^{\omega\rho},
\label{IL}
\end{eqnarray}
where $K_\nu(z)$ is the modified Bessel function. In the left overlapping region $I_L\cap II$ where
$-\omega\rho<<1$, the solutions
(\ref{IL}) can be approximated by
\begin{eqnarray}
u_1&\simeq& 1\nonumber\\
u_2&\simeq& \frac{3}{\omega^2\rho^2}\nonumber\\
u_3&\simeq& -\frac{1}{\omega\rho}\nonumber\\
u_3'&\simeq& 1,
\label{ILII}
\end{eqnarray}
matching the power law behavior dictated by
(\ref{DI_II}) and (\ref{DI_II_L}).
It follows from (\ref{DI_II}) and (\ref{DI_II_R}) in the right
overlapping region that
\begin{eqnarray}
u_1&\simeq& A_1+B_1\rho\nonumber\\
u_2&\simeq& \frac{1}{\omega^2}\left(\frac{A_2}{\rho^2}+B_2\rho^3\right)\nonumber\\
u_3&\simeq& \frac{1}{\omega}(A_3+B_3\rho)\nonumber\\
u_3'&\simeq& \frac{1}{\omega}(\frac{A_3^\prime}{\rho}+B_3^\prime\rho^2),
\label{IRII}
\end{eqnarray}
where the coefficients A's and B's are determined by
(\ref{ILII}) and the solutions covering the entire domain II, with $D$'s given by (\ref{DII})
and are {\it independent} of $\omega$.
Coming to $I_R$, the solutions take the approximate forms
\begin{eqnarray}
u_1&\simeq&2(C_1\sinh\omega\rho+\bar C_1e^{-\omega\rho})\nonumber\\
u_2&\simeq&\sqrt{2\pi\omega\rho}[C_2I_{\frac{5}{2}}(\omega\rho)
+\bar C_2K_{\frac{5}{2}}(\omega\rho)]\nonumber\\
u_3&\simeq& 2(C_3\sinh\omega\rho+\bar C_3e^{-\omega\rho}) \nonumber\\
u_3'&\simeq&\sqrt{2\pi\omega\rho}[C_3I_{\frac{3}{2}}(\omega\rho)+\bar C_3^\prime K_{\frac{3}{2}}(\omega\rho)],
\label{IR}
\end{eqnarray}
following (\ref{D12I}) and (\ref{D3IR}). All these solutions grows exponentially as
$\rho\to\infty$. The reason why $u_3(\rho)$ and $u_3'(\rho)$ share the same coefficient of exponentially growing follows from
the fact that $u_3^\prime(-\rho)$ is the kernel of $D_3$ and its Wronskian with $u_3(\rho)$ is independent of $\rho$.
Matching the approximate solutions of (\ref{IRII}) and (\ref{IR}),
we find that
\begin{eqnarray}
C_1&\simeq&\frac{B_1}{2\omega}\nonumber\\
C_2&\simeq&\frac{15B_2}{2\omega^5}\nonumber\\
C_3&\simeq&\frac{B_3}{2\omega^2}
\label{lowomega}
\end{eqnarray}
and
\begin{equation}
\bar C_1(\omega)=O(1), \qquad \bar C_2(\omega)=O(1),
\qquad \bar C_3=O(\omega^{-1}), \qquad C_3^\prime(\omega)=O(\omega)
\label{orders}
\end{equation}
for $\omega<<1$.

The coefficients $B$'s can be figured out analytically. It is more convenient
to work with the old variable $r$ defined by the transformation (\ref{rho}). A large $r$ corresponds
to a large magnitude of $\rho$ according to
\begin{equation}
\rho\simeq\pm\frac{\sqrt{2}r}{r_h}
\label{overlap}
\end{equation}
with positive(negative) sign for $\rho>0$($\rho<0$). Near the origin $\rho=0$, we have
\begin{equation}
\sqrt{1-\frac{r_h^4}{r^4}}\simeq\pm\frac{\rho}{\sqrt{2}}
\label{origin}
\end{equation}
as $\rho\to 0^\pm$. In terms of $r$-coordinates, we have
\begin{eqnarray}
D_1&=&-\frac{r_h^2}{2}\sqrt{f}\frac{d}{dr}\sqrt{f}\frac{d}{dr}+\frac{r_h^6}{r^6}\nonumber\\
D_2&=&-\frac{r_h^2}{2}\sqrt{f}\frac{d}{dr}\sqrt{f}\frac{d}{dr}+\frac{r_h^2}{r^2}\left(3+\frac{r_h^4}{r^4}\right),
\label{D1}
\end{eqnarray}
and the equations $D_1u_1=0$ and $D_2u_2=0$ can be reduced to hypergeometric equations with
the solutions that matches the powers of (\ref{ILII})
\begin{eqnarray}
u_1&=&F\left(\frac{1}{2},\frac{1}{4};\frac{5}{4};z\right)\nonumber\\
u_2&=&\frac{3}{2\omega^2}z^{\frac{1}{2}}F\left(1,\frac{3}{4};\frac{9}{4};z\right).
\label{sol_L}
\end{eqnarray}
for $\rho<0$, where $z=\frac{r_h^4}{r^4}$ and the asymptotic form (\ref{overlap}) has been used for matching.
It follows from (\ref{origin}) and the formula
\begin{eqnarray}
F(a,b;c;z)&=&\frac{\Gamma(c)\Gamma(c-a-b)}{\Gamma(c-a)\Gamma(c-b)}F(a,b;a+b-c+1;1-z)\nonumber\\
&+&\frac{\Gamma(c)\Gamma(a+b-c)}{\Gamma(a)\Gamma(b)}(1-z)^{c-a-b}F(c-a,c-b;c-a-b+1;1-z)
\end{eqnarray}
that the analytic continuation of (\ref{sol_L}) to $\rho>0$ reads
\begin{eqnarray}
u_1&=&-F\left(\frac{1}{2},\frac{1}{4};\frac{5}{4};z\right)
+\frac{1}{2\sqrt{2\pi}}\Gamma^2\left(\frac{1}{4}\right)z^{-\frac{1}{4}}\nonumber\\
u_2&=&\frac{3}{2\omega^2}z^{\frac{1}{2}}F\left(1,\frac{3}{4};\frac{9}{4};z\right)
+\frac{15}{8\sqrt{2\pi}\omega^2}\Gamma^2\left(\frac{1}{4}\right)
z^{-\frac{3}{4}}F\left(-\frac{1}{4},-\frac{1}{2};-\frac{1}{4};z\right).
\label{sol_R}
\end{eqnarray}
Matching (\ref{sol_R}) to (\ref{IRII}), we find that
\begin{equation}
B_1=\frac{1}{4\sqrt{\pi}}\Gamma^2\left(\frac{1}{4}\right)\qquad
B_2=\frac{15}{32\sqrt{\pi}}\Gamma^2\left(\frac{1}{4}\right),
\end{equation}
which yields the small $\omega$ behavior
\begin{equation}
C_1\simeq\frac{\Gamma^2\left(\frac{1}{4}\right)}{8\sqrt{\pi}\omega}\qquad
C_2\simeq\frac{225\Gamma^2\left(\frac{1}{4}\right)}{64\sqrt{\pi}\omega^5}
\end{equation}
following from (\ref{lowomega}).

As to $C_3$, we notice that $D_3$ of (\ref{DII}) factorizes as
\begin{equation}
D_3=\left(\frac{d}{d\rho}-e^\phi\right)\left(\frac{d}{d\rho}+e^\phi\right).
\end{equation}
The solution of the 1st order equation
\begin{equation}
\left(\frac{d}{d\rho}+e^\phi\right)\psi=0
\label{1storder}
\end{equation}
also solves the 2nd order equation $D_3\psi=0$. In terms of the $r$-coordinate,
eq.(\ref{1storder}) takes the explicit form
\begin{eqnarray}
\left(\sqrt{f}\frac{d}{dr}\pm\frac{1}{r}\right)\psi=0
\end{eqnarray}
with the upper(lower) sign for negative(positive) $\rho$. Its solution
\begin{eqnarray}
\psi=\frac{1}{\omega}\left(\frac{r^2}{r_h^2}+
\sqrt{\frac{r^4}{r_h^4}-1}\right)^{\mp\frac{1}{2}}
\end{eqnarray}
matches the power behavior of $u_3$ in the left overlapping region, (\ref{ILII}) and joins smoothly
at $\rho=0$. Therefore we identify $u_3=\psi$. Matching $u_3$ to the behavior (\ref{IRII}),
we obtain $B_3=1$ and then
\begin{equation}
C_3\simeq\frac{1}{2\omega^2}
\end{equation}
as $\omega\to 0$.

\section {The limt $\lim_{\omega\to 0}b(\omega)$}

As $\omega\to 0$, we may find introduce $\rho_0$ such that $\rho_0>>1$ but
$|\omega\rho_0|<<1$ and divide the domain of the double integrals for $b(\omega)$ into
the following six parts:
\begin{eqnarray}
&&I: -\rho_0<\rho<\rho_0,\qquad -\rho_0<\rho'<\rho_0, \qquad \rho'<\rho;\nonumber\\
&&II: \rho_0<\rho'<\rho;\nonumber\\
&&III: \rho>\rho_0, \qquad -\rho_0<\rho'<\rho_0;\nonumber\\
&&IV: \rho>\rho_0, \qquad \rho'<-\rho_0;\nonumber\\
&&V: -\rho_0<\rho<\rho_0, \qquad \rho'<-\rho_0;\nonumber\\
&&VI: \rho'<\rho<-\rho_0.
\end{eqnarray}
Correspondingly, we write
\begin{equation}
b(\omega)=b_I(\omega)+b_{II}(\omega)+b_{III}(\omega)+b_{IV}(\omega)+b_V(\omega)+b_{VI}(\omega).
\end{equation}
If $-\rho_0<\rho<\rho_0$, the solutions $u_1$, $u_2$, $u_3$ and $v_3$ of the integrand may be approximated
by the kernels of (\ref{DII}). Otherwise, the asymptotic forms (\ref{IL}) or (\ref{IR}) are substituted
for $\rho<-\rho_0$ or $\rho>\rho_0$. It follows from the approximate solutions in the appendix C that
\begin{eqnarray}
&&\int_{-\rho_0}^{\rho_0} d\rho e^{\phi(\rho)} u_{1\omega}
(-\rho)u_{2\omega}(-\rho)\int_{-\rho_0}^{\rho_0} d\rho' e^{\phi(\rho')} u_{1\omega}
(\rho')u_{2\omega}(\rho')=O(\omega^{-4})\nonumber\\
&&\int_{-\rho_0}^{\rho_0} d\rho e^{\phi(\rho)} v_{3\omega}^2
(\rho)\int_{-\rho_0}^{\rho_0} d\rho' e^{\phi(\rho')} u_{3\omega}^2(\rho')=O(\omega^{-2})\nonumber\\
&&\int_{\rho_0}^\infty d\rho e^{\phi(\rho)} u_{1\omega}
(-\rho)u_{2\omega}(-\rho)\int_{-\rho_0}^{\rho_0} d\rho' e^{\phi(\rho')} u_{1\omega}
(\rho')u_{2\omega}(\rho')=O(\omega^{-4})\nonumber\\
&&\int_{\rho_0}^\infty d\rho e^{\phi(\rho)} v_{3\omega}^2
(\rho)\int_{-\rho_0}^{\rho_0} d\rho' e^{\phi(\rho')} u_{3\omega}^2(\rho')=O(\omega^{-2}\ln\omega)\nonumber\\
&&\int_{\rho_0}^\infty d\rho e^{\phi(\rho)} u_{1\omega}
(-\rho)u_{2\omega}(-\rho)\int_{-\infty}^{-\rho_0} d\rho' e^{\phi(\rho')} u_{1\omega}
(\rho')u_{2\omega}(\rho')=O(\omega^{-4})\nonumber\\
&&\int_{\rho_0}^\infty d\rho e^{\phi(\rho)} v_{3\omega}^2
(\rho)\int_{-\infty}^{-\rho_0} d\rho' e^{\phi(\rho')} u_{3\omega}^2(\rho')=O(\omega^{-2}\ln\omega)\nonumber\\
&&\int_{-\rho_0}^{\rho_0} d\rho e^{\phi(\rho)} u_{1\omega}
(-\rho)u_{2\omega}(-\rho)\int_{-\infty}^{-\rho_0} d\rho' e^{\phi(\rho')} u_{1\omega}
(\rho')u_{2\omega}(\rho')=O(\omega^{-4})\nonumber\\
&&\int_{\rho_0}^\infty d\rho e^{\phi(\rho)} v_{3\omega}^2
(\rho)\int_{-\infty}^{-\rho_0} d\rho' e^{\phi(\rho')} u_{3\omega}^2(\rho')=O(\omega^{-2})\nonumber\\
\end{eqnarray}
Upon multiplication with $C's$, we find that
\begin{equation}
\lim_{\omega\to 0}b_I(\omega)=\lim_{\omega\to 0}b_{III}(\omega)=\lim_{\omega\to 0}b_{IV}(\omega)
=\lim_{\omega\to 0}b_V(\omega)=0,
\end{equation}
because of (\ref{csmall}). As to $b_{II}(\omega)$ and $b_{VI}(\omega)$, we have
\begin{eqnarray}
&&\int_{\rho_0}^\infty d\rho e^{\phi(\rho)} u_{1\omega}
(-\rho)u_{2\omega}(-\rho)\int_{\rho_0}^\rho d\rho' e^{\phi(\rho')} u_{1\omega}
(\rho')u_{2\omega}(\rho')\nonumber\\
&=&\int_{-\infty}^{-\rho_0} d\rho e^{\phi(\rho)} u_{1\omega}
(-\rho)u_{2\omega}(-\rho)\int_{-\infty}^\rho d\rho' e^{\phi(\rho')} u_{1\omega}
(\rho')u_{2\omega}(\rho')\nonumber\\
&=&4\int_{\omega\rho_0}^\infty\frac{dx}{\sqrt{x}}e^{-x}K_{\frac{5}{2}}(x)
\int_{\omega\rho_0}^x\frac{dy}{\sqrt{y}}[C_1(\omega)\sinh y+\bar{C_1}(\omega)e^{-y}]
[C_2(\omega)I_{\frac{5}{2}}(y)+\bar{C_2}(\omega)K_{\frac{5}{2}}(y)]\nonumber\\
&&\int_{\rho_0}^\infty d\rho e^{\phi(\rho)} v_{3\omega}^2(\rho)
\int_{\rho_0}^\rho d\rho' e^{\phi(\rho')} u_{3\omega}^2(\rho')\nonumber\\
&=&4\int_{\omega\rho_0}^\infty\frac{dx}{x}e^{-2x}\int_{\omega_0\rho_0}^x\frac{dy}{y}
[C_3(\omega)\sinh y+\bar{C_3}(\omega)e^{-y}]^2
\int_{-\infty}^{-\rho_0} d\rho e^{\phi(\rho)} v_{3\omega}^2(\rho)
\int_{-\infty}^\rho d\rho' e^{\phi(\rho')} u_{3\omega}^2(\rho')\nonumber\\
&=&4\int_{\omega\rho_0}^\infty dxK_{\frac{3}{2}}^2(x)\int_{\omega_0\rho_0}^xdy
[C_3(\omega)I_{\frac{3}{2}}(y)+C_3^{(2)}(\omega)K_{\frac{3}{2}}(y)]^2.
\end{eqnarray}
It follows from (\ref{orders}) that
\begin{equation}
\lim_{\omega\to 0}b_{II}(\omega)=-16\int_0^\infty\frac{dx}{\sqrt{x}}e^{-x}K_{\frac{5}{2}}(x)
\int_0^x\frac{dy}{\sqrt{y}}I_{\frac{5}{2}}(y)
+16\int_0^\infty\frac{dx}{x}e^{-2x}\int_0^x\frac{dy}{y}\sinh^2y
\end{equation}
and
\begin{equation}
\lim_{\omega\to 0}b_{VI}(\omega)=-16\int_0^\infty\frac{dx}{\sqrt{x}}e^{-x}K_{\frac{5}{2}}(x)
\int_0^x\frac{dy}{\sqrt{y}}I_{\frac{5}{2}}(y)
+16\int_0^\infty dxK_{\frac{3}{2}}(x)\int_0^x dyI_{\frac{3}{2}}(y).
\end{equation}
Therefore
\begin{eqnarray}
\lim_{\omega\to 0}b(\omega)&=&-32\int_0^\infty\frac{dx}{\sqrt{x}}e^{-x}K_{\frac{5}{2}}(x)
\int_0^x\frac{dy}{\sqrt{y}}I_{\frac{5}{2}}(y)
+16\int_0^\infty\frac{dx}{x}e^{-2x}\int_0^x\frac{dy}{y}\sinh^2y\nonumber\\
&&+16\int_0^\infty dxK_{\frac{3}{2}}(x)\int_0^x dyI_{\frac{3}{2}}(y)\simeq 2.
\end{eqnarray}


\end{document}